\documentclass{aa} 
\usepackage{graphicx}
\usepackage{longtable}
\usepackage{txfonts}
\usepackage{natbib}
\usepackage{lscape}                                
\usepackage{color}
\usepackage{upgreek}
\newcommand{\hii} {H\,{\sc ii}}
\newcommand{\Teff} {T$_{\rm eff}$}
\newcommand{\grav} {log\,{\em g}}
\newcommand{\vsini} {{\it v}\,sin\,$i$}
\newcommand{\micro} {$\zeta_{\rm t}$}
\newcommand{\kms} {km~s$^{-1}$}
\newcommand{\idl} {{\sc IDL}}

\newcommand{\solu}[2]{#1\,$\pm$\,#2}
\newcommand{\ioni}[2]{{#1\,\sc{#2}}}

\newcommand{\fastwind}{{\sc FASTWIND}}
\newcommand{\iraf}{{\sc IRAF}}
\newcommand{\pyneb}{{\sc PyNeb}}

\newcommand{\nordic}{NOT}
\newcommand{\herschel}{WHT}

\newcommand{\isis}{ISIS}

\newcommand{\ha}{{H$\alpha$}}
\newcommand{\hb}{{H$\beta$}}

\newcommand{\Te}{$T_{\rm e}$}
\newcommand{\Ne}{$n_{\rm e}$}
\newcommand{\te}{$T_{\rm e}$}

\newcommand{\foii}{[O~{\sc ii}]}
\newcommand{\fsii}{[S~{\sc ii}]}
\newcommand{\fsiii}{[S~{\sc iii}]}

\newcommand{\fnii}{[N~{\sc ii}]}

\newcommand{\hi}{H\,{\sc i}}
\newcommand{\hei}{He~{\sc i}}

\begin{document}
%
\title{The Cocoon Nebula and its ionizing star: 
 do stellar and nebular abundances agree?\thanks{Based 
on observations made with the William Herschel Telescope operated by the Isaac Newton Group and with  the Nordic Optical Telescope, operated by the Nordic Optical Telescope Scientific Association. Both telescopes are at the Observatorio del Roque de los Muchachos, La Palma, Spain, of the Instituto de Astrof\'{\i}sica de Canarias.}}

\author{J. Garc\'ia-Rojas\inst{1,2}, S. Sim\'on-D\'{\i}az\inst{1,2} \& C. Esteban\inst{1,2}}

\institute{Instituto de Astrof\'\i sica de Canarias, E-38200 
           La Laguna, Tenerife, Spain
	   \and
           Universidad de La Laguna, Dept. Astrof\'isica,  E-38204 
           La Laguna, Tenerife, Spain.
	   }
\offprints{jogarcia@iac.es}

\date{Submitted 23 July 2014 /Accepted 29 September 2014}
\titlerunning{Cocoon}
\authorrunning{J. Garc\'{\i}a-Rojas et al.}

%
\abstract
{Main sequence massive stars embedded in an \hii\ region should have the same chemical abundances as the surrounding nebular gas+dust.  The Cocoon nebula (IC\,5146), a close-by Galactic \hii\ region ionized by a narrow line B0.5~V single star (BD+46\,3474), is an ideal target to perform a detailed comparison of nebular and stellar abundances in the same Galactic \hii\ region.
}
{We investigate the chemical content of oxygen and other elements in the Cocoon nebula from two different points of view: an empirical analysis of the nebular spectrum and a detailed spectroscopic analysis of the associated early B-type star using state-of-the-art stellar atmosphere modeling. By comparing the stellar and nebular abundances, we aim to indirectly address the long-standing problem of the discrepancy found between abundances obtained from collisionally excited lines and optical recombination lines in photoionized nebulae.
}
{We collect long-slit spatially resolved spectroscopy of the Cocoon nebula and a high resolution optical spectrum of the ionizing star. Standard nebular techniques along with updated atomic data are used to compute the physical conditions and gaseous abundances of O, N and S in 8 apertures extracted across a semidiameter of the nebula. We perform a self-consistent spectroscopic abundance analysis of BD+46~3474 based on the atmosphere code FASTWIND to determine the stellar parameters and Si, O, and N abundances.
}
{
The Cocoon nebula and its ionizing star, located at a distance of 800$\pm$80 pc, have a very similar chemical composition as the Orion nebula and other B-type stars in the solar vicinity. This result agrees with the high degree of homogeneity of the present-day composition of the solar neighbourhood (up to 1.5 Kpc from the Sun) as derived from the study of the local cold-gas ISM.
The comparison of stellar and nebular collisionally excited line abundances in the Cocoon nebula indicates that O and N gas+dust nebular values are in better agreement with  stellar ones assuming small temperature fluctuations, of the order of those found in the Orion nebula ($t^2=0.022$). For S, the behaviour is somewhat puzzling, reaching to different conclusions depending on the atomic data set used.
}
{}
%
\maketitle
%
%
\section{Introduction}
\label{section1}
%

In nebular astrophysics, the oxygen abundance is the most widely used proxy of metallicity from  the Milky Way to far-distant galaxies. A precise knowledge of its abundance, as well as of nitrogen, carbon, $\alpha$-elements and iron-peak elements and their ratios at different redshifts are crucial to understand the nucleosynthesis processes in the stars and the chemical evolution history of the Universe \citep{henryetal00, chiappinietal03, carigietal05}. Uncertainties in the knowledge of the oxygen abundance (i.e., metallicity) have important implications in several topics of modern Astrophysics such as the luminosity- and mass-metallicity relations for local and high-redshift star-forming galaxies \citep{tremontietal04}, the calibrations of strong line methods for deriving the abundance 
scale of extragalactic {\hii} regions and star-forming galaxies at different redshifts \citep{peimbertetal07, penaguerreroetal12b}, or the determination of the primordial helium abundance \citep{peimbert08}.

Ionized nebulae have been always claimed to be the most reliable
and straightforward astrophysical objects to determine abundances from close-by to large distances 
in the Universe. However, they are not free from some difficulties. One of the most longstanding problems is the dichotomy systematically found between the nebular abundances provided by the standard method, based on the analysis of intensity ratios of collisionally excited lines (CELs, which strongly depend on the assumed physical conditions in the nebula) and the abundances given by the faint optical recombination lines (ORLs, which are almost insensitive to the adopted physical conditions).

Already pointed out in the pioneering work by \citet{wyse42} in the planetary nebula NGC\,7009 more than seventy years ago, the observational evidence of CEL and ORL providing discrepant results has increased significantly in the last decade,
both using Galactic {\hii} regions data \citep[][and references therein]{estebanetal05, estebanetal13, garciarojasesteban07} and extragalactic {\hii} regions data \citep{estebanetal02, estebanetal09, estebanetal14, apeimbert03, lopezsanchezetal07, penaguerreroetal12b}.
In particular, these authors have found that 
the O$^{2+}$/H$^+$ ratio computed from \ion{O}{ii} ORLs gives sistematically higher values than that obtained from [\ion{O}{iii}] CELs by a factor ranging $\sim$1.2$-$2.2. Similar discrepancies have been reported for other ions 
for which abundances can be determing using both CELs and ORLs: 
C$^{2+}$, Ne$^{2+}$ and O$^+$ \citep[see][]{garciarojasesteban07}. The origin of this
discrepancy is still unkonwn and has been subject of debate for many years
\citep[see e.g.,][and references therein for the most recent literature on the subject]{tsamispequignot05, stasinskaetal07, mesadelgadoetal09, mesadelgadoetal12, tsamisetal11, nichollsetal12, nichollsetal13, peimbertpeimbert13}. 

The comparison of nebular abundances with those resulting from the spectroscopic analysis of associated blue massive stars (especially B-type stars in the main sequence and BA Supergiants) is another way to shed some light in the nebular abundance conundrum, particularly, in the absence of nebular ORLs. These stars have been proved to be powerful alternative tools to derive the present-day chemical composition of the interstellar material in the Galactic regions where they are located (similarly to \hii\ regions).
Following this approach, several authors have compared
galactic radial abundance gradients obtained from massive stars and {\hii} regions 
in nearby spiral galaxies
\citep[see][for NGC\,300, M\,31 and M\,33, respectively]{bresolinetal09, trundleetal02, uetal09}. 
When examined together, the outcome of these studies is, however, not completely conclusive. For example, while
a total agreement between nebular and stellar abundances is observed in NGC\,300 \citep{bresolinetal09}, a remarkable discrepancy is found in M\,31 \citep{trundleetal02}.  However, the 
comparison of nebular and stellar abundances in most of these studies are performed in a
global way, while there still exists the possibility of local or azimuthal variations
of abundances in the studied galaxies hampering a meaningful comparison of abundances. In addition, in several cases, nebular abundances could be only obtained by means of indirect (strong-line) methods and not directly from CELs/ORLs.   

To avoid some of these limitations one should concentrate on the study of
\hii\ regions and close-by blue massive stars for which we are confident that have been
formed and evolved in the same environment (and hence share the same chemical composition).
In addition, a hadful set of ``technical'' conditions must be taken into account:
a) the {\hii} region must be bright enough in order to detect the relatively faint auroral lines to determine {\te} and, if possible, a handful set of ORLs, b) the spectra of the stars must show a statistically meaningful set of non-blended metallic lines
(stars with spectral types in the range O9-B2 and low projected rotational velocities, \vsini\ $<$70 km s$^{-1}$, are the most suitable targets for this study), c) stars must not belong to binary/multiple systems.
There are not many
Galactic {\hii} regions that fulfill all these conditions. Of course, the keystone of all nebular studies, the Orion nebula (M~42), is one of them\footnote{Among the few others we can cite M~16 (Eagle nebula), NGC3372 (Carina nebula), or IC5146 (Cocoon nebula, studied in this paper)}. The comparison of nebular and stellar abundances in the Orion star forming region (as derived from the analysis of the spectra of M~42 and a handful number of early B-type stars in the region\footnote{The nebular abundance analysis was performed by \cite{estebanetal04}, and revisited by \cite{simondiazstasinska11}; the stellar abundance analysis was performed by \cite{simondiaz10} and \cite{nievasimondiaz11}.}) has been recently reviewed by \citet{simondiazstasinska11}. The main conclusion of this study \citep[which has been actually present in the literature in the last two decades, see e.g.,][]{estebanetal98, estebanetal04} is that oxygen (gas+dust) nebular abundance based on ORLs agrees much better with the stellar abundances than the one derived from CELs.
If a similar thoughtful analysis of a larger sample of targets (including the analysis of other elements and considering different metallicity environments) confirm this result, this will have
important implications for several fields of modern Astrophysics and the way the abundance scale in the Universe is defined.


The Cocoon nebula (also known as IC\,5146, Caldwell 19 and Sh\,2-125) is an emission nebula located in the constelation of Cygnus. Its distance is somewhat uncertain and has been fixed between 950$\pm$80 pc \citep{harveyetal08} and 1200$\pm$180 pc \citep{herbigdahm02}. This nebula is an ideal target to perform a detailed comparison of nebular and stellar abundances in the same Galactic region since most of the ``technical'' conditions quoted above are fulfilled\footnote{We do not detect ORLs in the spectrum of the Cocoon nebula, but the discussion can be done in other terms (see Sect.~\ref{section5}).}. The nebula is ionized by a single B0.5 V star $-$BD46\,3474, with low projected rotational velocity$-$ and is bright enough to compute physical conditions and chemical abundances from its nebular emission line spectrum. In addition, the ionization degree of the nebula is low due to the relatively low effective temperature of the central star and then, the most relevant elements are only once ionized and no ionization correction factors (ICF) are needed for determining total abundances of some key-elements. 

In this paper we perform a detailed spectroscopic abundance analysis of a set of 
long-slit spatially resolved intermediate-resolution spectra of the Cocoon nebula and a high-resolution optical spectrum of the ionizing star. The derived nebular CEL and stellar abundances of O, N, and S are hence compared. The paper is structured as follows. The observational data\,set is presented in Sect. \ref{section2}. The nebular physical conditions and abundances are determined in Sect. \ref{section3}. 
A self-consistent spectroscopic abundance analysis of BD+46\,3474 is performed in Sect. \ref{section4}. 
The discussion of results and the main conclusions from our study
are presented in Sect.~\ref{section5} and Sect.~\ref{section6}, respectively.

\section{The observational data\,set}\label{section2}
%

\subsection{Nebular spectroscopy}
\label{section21}

%
\begin{figure}[!ht]
\centering
\includegraphics[width=8.5cm,angle=0]{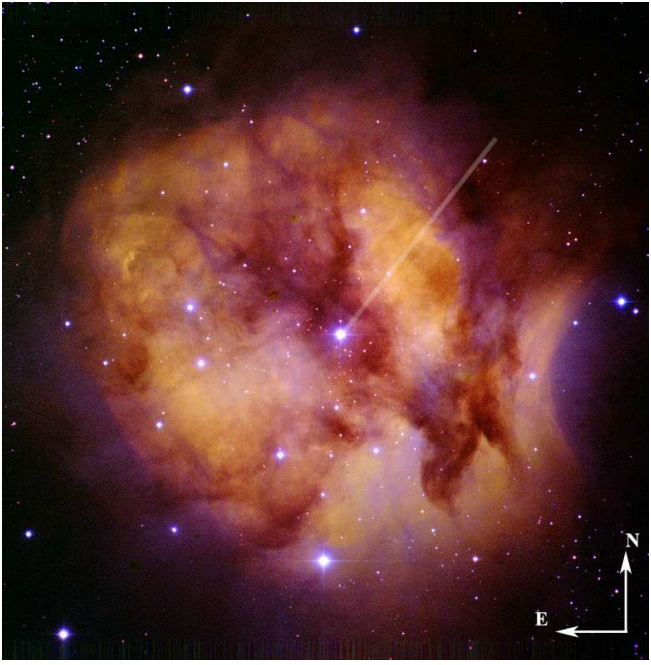}
\caption{Color composite (RGB) image of the Cocoon Nebula obtained from a 
combination of wide and narrow band images taken with the Wide Field Camera (WFC) 
at the 2.5m Isaac Newton Telescope (only part of the CCD~\#4 is shown) at the Roque de los Muchachos Observatory. The following color 
code was used: {\ha} (red), {\hb} (green), B (blue) (credit of the image by \'Angel R. L\'opez-S\'anchez, Australian Astronomical Observatory).
The field-of-view of the image is $\sim$\,11$'$ $\times$\,11$'$. The position of the slit used to obtain the nebular ISIS@WHT spectroscopy is shown. The PA 
was 320$^{\circ}$. Slit width is not at scale. 
}
\label{f1}
\end{figure}
%

A long-slit, intermediate-resolution spectrum of the Cocoon nebula was obtained on 
2011 November 23 with the Intermediate dispersion Spectrograph and Imaging 
System (\isis) spectrograph attached to the 4.2m William Herschel 
Telescope (\herschel) at Roque de los Muchachos observatory in La Palma (Canary Islands, Spain). 
The 3.7$'$\,$\times$\,1.02$''$ slit was located to the north-west of BD+46\,3474, (PA=320$^{\circ}$, see Fig.~\ref{f1}). Two different CCDs were used at the blue and red arms of 
the spectrograph: an EEV CCD with a configuration 4096 $\times$ 2048 pixels at 
13.5 $\mu$m, and a RedPlus CCD with 4096 $\times$ 2048 pixels 
at 15.0 $\mu$m, respectively. The dichroic prism used to separate the blue and red beams was set at 5300 \AA\ . The gratings R600B and R316R were used for the blue and 
red observations, respectively. These gratings give a reciprocal dispersion of 33 \AA\ mm$^{-1}$ in the blue and 62 \AA\ mm$^{-1}$ in the red, effective spectral resolutions of 2.2 and 3.56 \AA\, and the spatial scales are 0.20$''$ pixel$^{-1}$ and 0.22$''$ pixel$^{-1}$, respectively. We used two different grating angles in order to cover all the optical wavelength range: a) blue spectra centered at 4298 \AA\ and red one centered at 6147 \AA\ , and b) blue spectra centered at 4298 \AA\ and red one centered at 8148 \AA\ .  
The blue spectra cover an unvignetted range from $\lambda\lambda$3600 to 5100 \AA\ 
and the red ones from $\lambda\lambda$5498 to 9199 \AA.  The seeing during the 
observations was $\sim$2.0$''$. The exposure times were 4$\times$1200\,s in the blue, 3$\times$1200\,s in the red ($\lambda_c$ = 6147 \AA\ ) and 1$\times$1200\,s in the far  red ($\lambda_c$ = 8148 \AA\ ) observations.

%
\begin{figure}[!ht]
\centering
\includegraphics[width=8.5cm,angle=0]{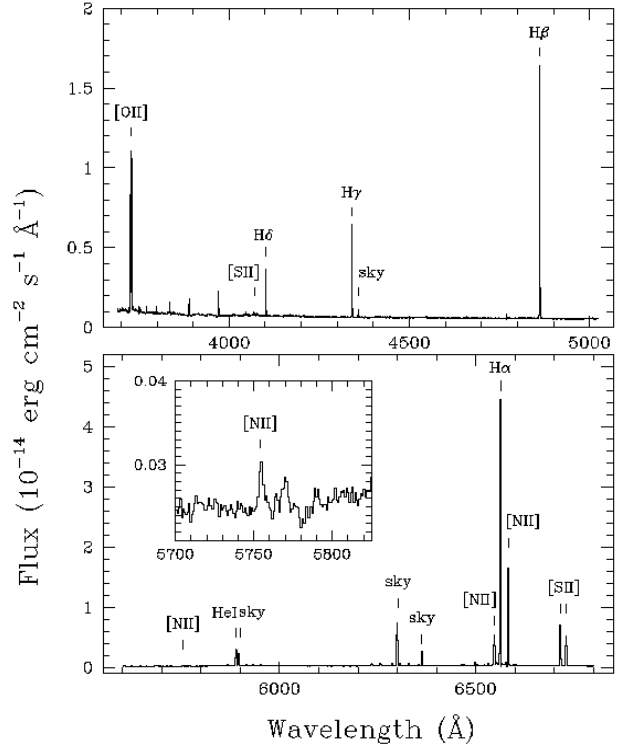}
\caption{Sections of the ISIS-WHT optical spectrum of the cocoon nebula (aperture \#2) with some lines of interest labelled. The inset shows a zoom of the zone around {\fnii} 5755 \AA\ . The mercury Hg~I 4358 \AA\  line and other night-sky features are also indicated. Sky emission could not be removed from the spectrum (see text).}
\label{f2}
\end{figure}
%

The spectra were wavelength calibrated with a CuNe+CuAr lamp. The correction 
for atmospheric extinction was performed using the average curve for continuous 
atmospheric extinction at Roque de los Muchachos Observatory. The absolute 
flux calibration was achieved by observations of the standard star 
BD+45\,4655. We used the \iraf\footnote{\iraf\ is distributed by National Optical Astronomical Observatory which is operated by Association of Universities for Research in Astronomy Inc., under cooperative agreement with NSF}\ TWODSPEC reduction 
package to perform bias correction, flat-fielding, cosmic-ray rejection, wavelength 
and flux calibration. We checked the relative flux calibration between the bluest and reddest wavelengths by calibrating the spectrophotometric standard by itself, finding a relative flux calibration uncertainty better than $\sim$5 \%. Fig.~\ref{f2} shows an illustrative example of the quality of our nebular spectroscopic observations, where the main nebular lines used in this study are indicated. The sky emission could not be removed because the nebular emission was present along the whole slit.

\subsection{Stellar spectroscopy}
\label{section22}

The spectroscopic observations\footnote{The FIES spectrum of BD+46\,3474 was obtained during one of the observing nights of the {\em IACOB spectroscopic database of Northern Galactic OB stars} program \citep{simondiazetal11b}.} of BD+46\,3474 were carried out with the FIES  
\citep{telting14} cross-dispersed, high-resolution echelle spectrograph attached to the 2.56m \nordic\ telescope at El Roque de los Muchachos observatory on 2012 September 10. The 
low-resolution mode ($R$ = 25000, $\delta\lambda$ = 0.03 \AA\ /pix) was selected, and the entire spectral range 3700$-$7100 \AA\ was covered without gaps in a single fixed setting. We took one single spectrum with an exposure time of 1200~s. The signal-to-noise ratio achieved was above 250.

\begin{figure*}[!ht]
\centering
\includegraphics[width=0.5\textwidth,angle=90]{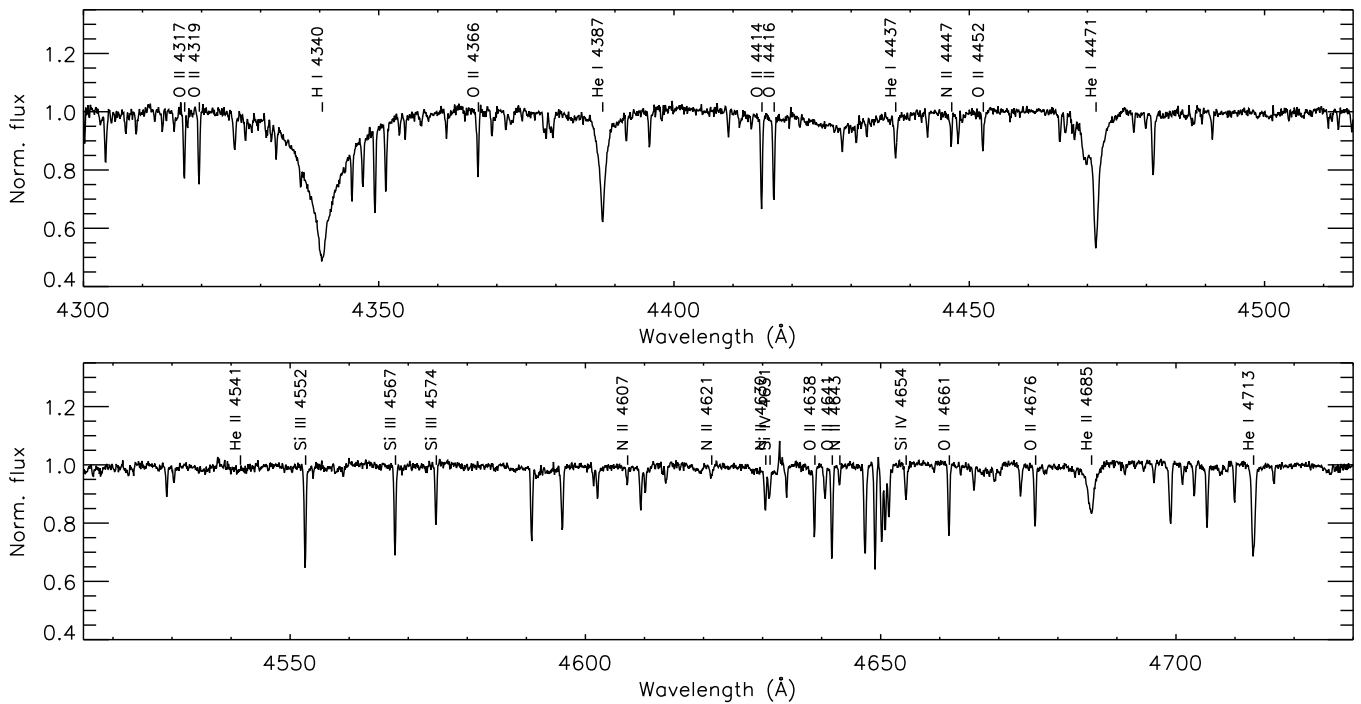}
\caption{Spectrum of BD+46\,3474 in the region between 4300 and 4730 \AA. In contrast
to the broad hydrogen Balmer and \ioni{He}{i} lines ($H_{\gamma}$ and 
\ioni{He}{i}~$\lambda\lambda$~4387, 4471 and 4713 in the plotted spectral window),
metal lines appear very narrow due to the low \vsini\ of the star. 
 Only the Si III-IV, O II and N II lines present in the plotted region and used for the spectroscopic abundance analyis are indicated in the figure (see text and Table 4).}
\label{f3}
\end{figure*}

The spectrum was reduced with the FIEStool\footnote{http://www.not.iac.es/instruments/fies/fiestool/FIEStool.html} 
software in advanced mode. The FIEStool pipeline provided a wavelength calibrated, blaze-corrected, order-merged spectrum of high quality. This spectrum was then normalized with our own developed \idl\ routines. A selected range of the spectrum of BD+46\,3474, where the main diagnostic lines used for the stellar parameter and abundance determination are indicated, is presented in Fig~\ref{f3}.

%
%

%
\section{Empirical analysis of the nebular spectra}
\label{section3}
%

\subsection{Aperture selection and line flux measurements}
\label{section31}

We obtained 1D spectra of zones of the nebula at different distances
from the central star by dividing the long-slit used for the \isis@\herschel\ nebular 
observations in 8 apertures within the limits of the nebula. 
Additionally we extracted a spectrum containing all the apertures to compare with the results obtained from the individual ones.

The size of the apertures, except the integrated one, was 22$''$. First aperture was centered 62.8$''$ away BD+46\,3474 in order to avoid strong dust scattered stellar continuum. A star about 78.3$''$ away BD+46\,3474 was also avoided. The positions of the apertures are summarized in Table \ref{t1}. 

The extraction of 1D spectra from each aperture was done using the \iraf\ 
task $apall$. The same zone and spatial coverage was considered in the blue 
and red spectroscopic ranges. 

We detected {\hi} and {\hei} optical recombination lines, along with 
collisionally excited lines (CELs) of several low ionized ions, such as {\foii}, 
{\fnii}, {\fsii} and {\fsiii}. Line fluxes were 
measured using the SPLOT routine of the \iraf\ package by integrating all the flux 
included in the line profile between two given limits and over a local continuum 
estimated by eye. 

Each emission line in the spectra was normalized to a particular {\hi} 
recombination line present in each wavelength interval: {\hb} for the blue range, {\ha} for 
the red spectra, and P11 for the far-red spectra, respectively. 

The reddening coefficient, $c$(H$\beta$) was obtained by fitting the observed H$\delta$/H$\beta$ and H$\gamma$/H$\beta$ line 
intensity ratios $-$the three lines lie in the same spectral range$-$ to the theoretical ones computed by \citet{storeyhummer95} for 
$T_e$\,=\,6500 K and $n_e$\,=\,100 cm$^{-3}$.

Finally, to produce a final homogeneous set of line intensity ratios, the red spectra were 
re-scaled to \hb\  applying the extinction correction and assuming the theoretical \ha/\hb\ and P11/\hb\ ratios: 
I(\ha)/I(\hb)\,=\,2.97 and I(P11)/I(\hb)\,=\,0.014 obtained assuming $T_e$\,=\,6500 K and $n_e$\,=\,100 cm$^{-3}$.

%
\begin{table*}
\centering \caption{Line fluxes$^{(1)}$ corrected from extinction (\hb\,=100), and
results from the empirical analysis$^{(2)}$ form CELs of the nebular spectra extrated from the 8 apertures and the integrated one. }
\label{t1}
\scriptsize
\begin{tabular}{c@{\hspace{2.8mm}}c@{\hspace{2.8mm}}c@{\hspace{2.8mm}}c@{\hspace{2.8mm}}c@{\hspace{2.8mm}}c@{\hspace{2.8mm}}c@{\hspace{2.8mm}}c@{\hspace{2.8mm}}c@{\hspace{2.8mm}}c@{\hspace{2.8mm}}c@{\hspace{2.8mm}}c@{\hspace{2.8mm}}}
\noalign{\hrule}
\noalign{\vskip3pt}
   &    &  & \multicolumn{9}{c}{Aperture} \\
\cline{4-12}
\noalign{\vskip3pt}
   &    &  &  A1  & A2  & A3  & A4  & A5  & A6  & A7 & A8  & Integrated \\
\cline{4-12}
\noalign{\vskip3pt}
\multicolumn{3}{r}{Center position$^{(3)}$ (arcsec)} &  62.8 & 93.8 & 115.8 & 137.8 & 159.8 & 181.8 & 203.8 & 225.8 & 144.3 \\
\multicolumn{3}{r}{Angular area (arcsec$^2$)} &  22 & 22 & 22 & 22 & 22 & 22 & 22 & 22 & 176 \\
\noalign{\vskip3pt}
\noalign{\hrule}
\noalign{\vskip3pt}
$\lambda$ (\AA) & Ion & Mult. & \multicolumn{7}{c}{$I$($\lambda$)/$I$(H$\beta$)}   \\
\noalign{\vskip3pt} \noalign{\hrule} \noalign{\vskip3pt} 
3726.03 & {\foii} & 1F & 74$\pm$6 & 72$\pm$2 & 73$\pm$3 & 80$\pm$2 & 94$\pm$5 & 93$\pm$7 & 107$\pm$5 & 129$\pm$18 & 80$\pm$4 \\
3728.82 & {\foii} & 1F & 103$\pm$7 & 102$\pm$3 & 102$\pm$3 & 112$\pm$3 & 130$\pm$7 & 129$\pm$9 & 149$\pm$6 & 172$\pm$22 & 111$\pm$5 \\
4068.60 & {\fsii} & 1F & 2.5$\pm$0.5 & 2.6$\pm$0.3 & 2.8$\pm$0.5 & 3.1$\pm$0.4 & 3.2$\pm$0.3 & 2.8$\pm$0.6 & 6$\pm$1 & $-$ & 2.8$\pm$0.4 \\
4076.35 & {\fsii} & 1F & 1.4$\pm$0.4 & 1.6$\pm$0.2 & 1.7$\pm$0.2 & 1.9$\pm$0.3 & 3.9$\pm$0.7 & 4$\pm$1 & 6$\pm$2 & $-$ & 2.3$\pm$0.3 \\
4101.74 & {\hi} & H$\delta$ & 25$\pm$1 & 25.3$\pm$0.7 & 25.8$\pm$0.7 & 25.6$\pm$0.7 & 26$\pm$2 & 27$\pm$2 & 26$\pm$2 & 24$\pm$5 & 26$\pm$1 \\
4340.47 & {\hi} & H$\gamma$ & 47$\pm$2 & 46$\pm$1 & 46$\pm$1 & 46$\pm$1 & 46$\pm$2 & 46$\pm$2 & 46$\pm$2 & 47$\pm$7 & 46$\pm$1 \\
4861.33 & {\hi} & H$\beta$ & 100$\pm$2 & 100$\pm$2 & 100$\pm$2 & 100$\pm$2 & 100$\pm$2 & 100$\pm$2 & 100$\pm$2 & 100$\pm$2 & 100$\pm$2 \\
5754.64 & {\fnii} & 3F & 0.51$\pm$0.07 & 0.53$\pm$0.05 & 0.62$\pm$0.08 & 0.6$\pm$0.1 & $-$ & $-$ & $-$ & $-$ & 0.5$\pm$0.1 \\ 
5875.64 & {\hei} & 11 & 2.1$\pm$0.2 & $-$ & $-$ & $-$ & $-$ & $-$ & $-$ & $-$ & 1.0$\pm$0.2 \\
6548.03 & {\fnii} & 1F &  36$\pm$2 & 35$\pm$1 & 38$\pm$1 & 37$\pm$1 & 39$\pm$3 & 41$\pm$2 & 39$\pm$1 & 42$\pm$4 & 37$\pm$2 \\
6562.82 & {\hi} & H$\alpha$ & 297$\pm$18 & 297$\pm$8 & 297$\pm$8 & 297$\pm$8 & 297$\pm$20 &  297$\pm$15 &  297$\pm$6 &  297$\pm$23 & 297$\pm$15 \\
6583.41 & {\fnii} & 1F &  111$\pm$7 & 108$\pm$3 & 109$\pm$3 & 113$\pm$3 & 115$\pm$8 & 122$\pm$6 & 124$\pm$3 & 121$\pm$9 & 111$\pm$ 6 \\
6678.15 & {\hei} & 46 & 0.44$\pm$0.06 & $-$ & $-$ & $-$ & $-$ & $-$ & $-$ & $-$ & $-$ \\
6716.47 & {\fsii} & 2F & 41$\pm$3 & 45$\pm$1 & 51$\pm$2 & 59$\pm$2 & 61$\pm$4 & 67$\pm$4 & 72$\pm$2 & 71$\pm$6 & 51$\pm$3 \\
6730.85 & {\fsii} & 2F & 30$\pm$2 & 33.2$\pm$0.9 & 38$\pm$1 & 43$\pm$1 & 44$\pm$3 & 49$\pm$3 & 52$\pm$1 & 50$\pm$4 & 37$\pm$2 \\
8862.79 & {\hi} & P11 & 1.4$\pm$0.2 & 1.4$\pm$0.1 & 1.4$\pm$0.1 & 1.4$\pm$0.2 & 1.4$\pm$0.5 & 1.4$\pm$0.4 & 1.4$\pm$0.3 & 1.4: & 1.4$\pm$0.2 \\ 
9014.91 & {\hi} & P10 & 1.1$\pm$0.3 & 1.2$\pm$0.1 & 1.2$\pm$0.1 & 1.1$\pm$0.2 & 1.0: & 0.6:& 0.4: & $-$ & 1.1$\pm$0.3 \\
9068.90 & {\fsiii} & 1F & 6$\pm$1& 6.5$\pm$0.4 & 4.5$\pm$0.3 & 3.5$\pm$0.3 & 2.9$\pm$0.8 & 2.5$\pm$0.8 & 1.9$\pm$0.6 & 1.7: & 4.4$\pm$0.6 \\
\noalign{\vskip3pt} \noalign{\hrule} \noalign{\vskip3pt}
\multicolumn{3}{r}{$c$(H$\beta$)} & 1.07$\pm$0.10 & 0.79$\pm$0.02 & 0.89$\pm$0.04 & 0.61$\pm$0.02 & 0.90$\pm$0.08  & 0.95$\pm$0.11 & 1.12$\pm$0.04 & 1.65$\pm$0.21 & 0.88$\pm$0.06 \\
\multicolumn{3}{r}{$F$(H$\beta$)$^{(4)}$} & 2.39$\pm$0.05 & 2.72$\pm$0.05 & 2.49$\pm$0.05 & 2.07$\pm$0.04 & 0.78$\pm$0.02 & 0.74$\pm$0.02 & 0.40$\pm$0.01 & 0.26$\pm$0.01 & 11.8$\pm$0.2 \\
\noalign{\vskip3pt} 
\multicolumn{3}{r}{\Ne(\foii)} & 63$^{+59}_{-63}$ & 50$\pm$28 & 56$\pm$35 & 53$\pm$25 &  59$^{+106}_{-59}$ & 58$^{+76}_{-58}$ & 41$\pm$29 & 28$^{+110}_{-28}$ & 60$\pm$43 \\
\multicolumn{3}{r}{\Ne(\fsii)} & 48$^{+80}_{-48}$ & 50$\pm$30 & 50$\pm$34 & 50$\pm$31 & 61$\pm$56 & 63$^{+69}_{-63}$ & 63$\pm$46 & 90$^{+149}_{-90}$ & 46$^{+60}_{-46}$ \\
\multicolumn{3}{r}{\Te(\fnii)} & 6850$\pm$280 & 7020$\pm$210 & 7220$\pm$280 & 7230$\pm$420 & $-$ & $-$ & $-$ & $-$ & 6900$\pm$490 \\
\multicolumn{3}{r}{\Te(\fsii)} & 8190$\pm$810 & 8030$\pm$430 & 7700$\pm$520 & 7650$\pm$390 & 9360$\pm$820 & 8800$\pm$1000 & 12000$\pm$2500 & $-$ & 8410$\pm$580 \\
\noalign{\vskip3pt} 
\multicolumn{3}{r}{O$^+$/H$^+$} & 8.63$\pm$0.06 & 8.55$\pm$0.05 & 8.48$\pm$0.06 & 8.52$\pm$0.10 &   $-$ & $-$ & $-$ & $-$ & 8.64$\pm$0.13 \\
\multicolumn{3}{r}{N$^+$/H$^+$} & 7.87$\pm$0.07 & 7.82$\pm$0.04 & 7.79$\pm$0.05 & 7.79$\pm$0.07 & $-$ & $-$ & $-$ & $-$ & 7.86$\pm$0.10 \\
\multicolumn{3}{r}{S$^+$/H$^+$} & 6.69$\pm$0.07 & 6.69$\pm$0.04 & 6.71$\pm$0.05 & 6.76$\pm$0.07 & $-$ & $-$ & $-$ & $-$ & 6.77$\pm$0.10 \\
\multicolumn{3}{r}{S$^{2+}$/H$^+$} & 6.30$\pm$0.12 & 6.17$\pm$0.07 & 6.10$\pm$0.08 & 5.99$\pm$0.11 & $-$ & $-$ & $-$ & $-$ & 6.15$\pm$0.15 \\
\noalign{\vskip3pt} 
\multicolumn{3}{r}{O/H} & 8.63$\pm$0.06 & 8.55$\pm$0.05 & 8.48$\pm$0.06 & 8.52$\pm$0.10 &  $-$ & $-$ & $-$ & $-$ & 8.64$\pm$0.13 \\
\multicolumn{3}{r}{N/H} & 7.87$\pm$0.07 & 7.82$\pm$0.04 & 7.79$\pm$0.05 & 7.79$\pm$0.07 & $-$ & $-$ & $-$ & $-$ & 7.86$\pm$0.10 \\
\multicolumn{3}{r}{S/H} & 6.84$\pm$0.06 & 6.81$\pm$0.04 & 6.80$\pm$0.04 & 6.83$\pm$0.06 & $-$ & $-$ & $-$ & $-$ & 6.86$\pm$0.08 \\
\noalign{\vskip3pt} \noalign{\hrule} \noalign{\vskip3pt}
\end{tabular} 
\begin{description}
\scriptsize
\item[] $^{(1)}$ The errors in the line fluxes only refer to uncertainties in the line measurements (see text). 
\item[] $^{(2)}$ {\Ne} in cm$^{-3}$; {\Te} in K; abundances in log(X$^{+i}$/H$^{+}$)+12.
\item[] $^{(3)}$ With respect to BD+46~3474.
\item[] $^{(4)}$ F(H$\beta$) in $\times$10$^{-14}$ erg cm$^{2}$ s$^{-1}$ and uncorrected for reddening.
\end{description}
\end{table*}

\subsection{Uncertainties}
\label{section32}

Several sources of uncertainties must be taken into account to obtain
the errors associated with the line intensity ratios. We estimated that 
the uncertainty
in the line intensity measurement due to the signal-to-noise
of the spectra and the placement of the local continuum is typically
$\sim$\,2\% for $F$($\lambda$)/$F$({\hb})$\geq$0.5, 
$\sim$\,5\% for 0.1$\leq$ $F$($\lambda$)/$F$({\hb}) $\leq$0.5, 
$\sim$\,10\% for 0.05$\leq$ $F$($\lambda$)/$F$({\hb}) $\leq$0.1, 
$\sim$\,20\% for 0.01$\leq$ $F$($\lambda$)/$F$({\hb})$\leq$0.05, 
$\sim$\,30\% for 0.005$\leq$ $F$($\lambda$)/$F$({\hb})$\leq$0.01, and 
$\sim$\,40\% for 0.001$\leq$ $F$($\lambda$)/$F$({\hb})$\leq$0.005.  
We did not consider those lines which are weaker than 0.001 $\times$ $F$({\hb}).
Note that uncertainties indicated in Table \ref{t1} only refer to this type 
of errors.
 
By comparing the resulting flux-calibrated spectra of our standard star 
with the corresponding tabulated flux, we could estimate that line ratio uncertainties 
associated to the flux calibration is 
$\sim$\,3\% when the wavelengths are separated by 500\,--\,1500 \AA\   and $\sim$\,5\% 
if they are separated by more than that. For the cases where 
the corresponding lines are separated by less than 500 \AA, the uncertainty 
in the line ratio due to uncertainties in the flux calibration is negligible.
 
The uncertainty associated to extinction correction was computed by error propagation.
Again, the contribution of this uncertainty to the total error is negligible when
line ratios of close-by lines are considered 
(e.g. [\ion{S}{ii}]\,$\lambda$6716/[\ion{S}{ii}]\,$\lambda$6730).
The final errors in the line intensity ratios used to derive the physical properties
of the nebula were computed by adding 
quadratically these three sources of uncertainty. 

\subsection{Physical conditions}
\label{section33}
%
Electron temperature (\te), and density (\Ne) of the ionized gas were
derived from classical CEL ratios, using \pyneb , a pyhton based code 
\citep{luridianaetal12} and the set of atomic data 
shown in Table~\ref{t1}. We computed {\Ne} from
the {\fsii} $\lambda$6717/$\lambda$6731 and {\foii} $\lambda$3729/$\lambda$3726
line ratios and {\te} from the nebular to auroral
{\fnii} $\lambda\lambda$(6548+84)/$\lambda$5754 line ratio and from {\fsii} $\lambda\lambda$(6717+31)/$\lambda$$\lambda$(4068+76).

The methodology used for the determination of the physical conditions was as follows:
we assumed a representative initial value of {\te} of 10000 K and compute the
electron densities. Then, the value of {\Ne} was used to compute 
{\te}({\fnii}) and/or {\te}({\fsii}) from the observed line ratios; for the four innermost apertures, we only assume {\te}({\fnii}) as valid (see below). 
We iterated until convergence to obtain the final values of {\Ne}  and {\te}. 
Uncertainties were computed by error propagation. The final {\Ne}({\foii}), 
{\Ne}({\fsii}), {\te}({\fnii}) and {\te}({\fsii}) estimations, along with their uncertainties
are indicated in Table~\ref{t1}. We do not rely on the determination of {\te}({\fsii}) for apertures 5$-$7, since it is much higher than those obtained in the other apertures, giving unreasonable low values for the chemical abundances of O, N and S when used. 

%
\begin{table}
\caption{Atomic data considered in the nebular abundance analysis.}
\label{t2}
\begin{tabular}{lll}
\noalign{\smallskip} \noalign{\smallskip} \noalign{\hrule} \noalign{\smallskip}
	&  \multicolumn{2}{c}{CELs} 			\\		       
Ion 	& Trans. Probabilities 	& Coll. strengths	\\
\noalign{\smallskip} \noalign{\hrule} \noalign{\smallskip}
N$^{+}$	& \citet{galavisetal97}	& \citet{tayal11}	\\
O$^{+}$	& \citet{zeippen82}	& \citet{pradhanetal06}	\\
S$^{+}$	&\citet{podobedovaetal09}& \citet{ramsbottometal96}\\
S$^{2+}$&\citet{podobedovaetal09}& \citet{galavisetal95}\\
\noalign{\smallskip} \noalign{\hrule} \noalign{\smallskip}
\end{tabular}
\end{table}

In general, densities derived from the {\foii} line ratio are very consistent with those derived from {\fsii} lines for all the apertures, and aditionally, they are also homogeneous for all the apertures.
On the other hand, {\te}'s derived from {\fnii} and {\fsii} line ratios present non negligible differences for a given aperture, especially in the inner aperture and in the total extraction. 
Given the high dependence of the value of {\te}({\fsii}) with the set of collisional strengths adopted and that S$^+$ zone do not strictly match with N$^+$ zone (S$^+$ is in the outer zones of the nebula, whereas N$^+$ is probably more evenly distributed along the nebula), we decided to adopt only {\te}({\fnii}) as representative of the whole nebula. 

%
\subsection{Chemical abundances}
\label{section34}
%

We have detected only two {\hei} lines in the innermost aperture. As hydrogen lines on this aperture could be affected by fluorescense effects (see below), we decided not computing the He$^+$ abundance that should be, in any case, a very low limit of the total He abundance due to the low ionization degree of the nebula.

Ionic abundances of N$^+$, O$^+$, S$^+$, and S$^{2+}$ were derived from CELs, using \pyneb . We
assumed a one zone scheme, in which we adopted {\te}({\fnii}) to compute N$^+$, O$^+$, S$^+$ and S$^{2+}$ abundances.
We assumed the average of {\Ne}({\fsii}) and {\Ne}({\foii}) as representative for each aperture. Due to the low \Teff\ of the ionizing star, we have not detected CELs of high ionization species, such as O$^{2+}$,  Ar$^{2+}$ or  Ne$^{2+}$. For the same reason, \ion{O}{ii} ORLs have not been detected in our spectrum. Additionally, neither the \ion{O}{i} ORLs in the 7771-7775 \AA\ range were detected in our spectrum because they are intrinsically faint and lie in a zone with strong sky emission; therefore, we could not compute the abundance discrepancies for O$^+$ nor for O$^{++}$ in this nebula (see Sect.~\ref{section1}).

We then derived total abundances of O, N and S for each aperture without using ionization correction factors (ICFs). We have to remark the importance of this because one of the main uncertainty sources in total abundances determinations in {\hii} regions come from the necessity of using ICFs (see more details in Sect. \ref{section5}).
The ionic abundances of  O$^{+}$, N$^{+}$, S$^{+}$, S$^{2+}$, along with the total abundances of O, N and S are shown in Table~\ref{t1}. 

From Table~\ref{t1} it can be seen that both physical conditions and total chemical abundances obtained from the innermost aperture, as well as from the integrated one are somewhat different than those obtained from apertures 2, 3 and 4, which are remarkably consistent one with each other. In principle one should not expect a variation in the total abundance obtained for a given element, especially if they were derived directly from ionic abundances, without using any ICF. \citet{ferland99} and \citet{luridianaetal09} described the importance of fluorescent excitation of Balmer lines due to continuum pumping in the hydrogen Lyman transitions by non-ionizing stellar continua. 
In particular,
\citet{luridianaetal09} performed a detailed description of this effect and its behaviour with the spectral type, luminosity class, and the distance to the star for enviroments where Lyman transitions are optically thick. In our case, this effect can be particularly important in aperture 1, which is the closest to the star. This 
effect depends on several parameters and it can only be adressed with a detailed photoionization model with an appropiate high resolution sampling of the non-ionizing spectrum of the stellar source \citep{luridianaetal09}. From a qualitative approach, differential fluorescent excitation of {\hi} lines can affect mainly the determination of the extinction coefficient, $c$(H$\beta$), which can be overestimated in the apertures closest to the ionizing star \citep[][ reported this effect in the inner apertures of a longslit study of M\,43]{simondiazetal11}; the overall effect would be i) an overestimation of the line fluxes bluer to H$\beta$ in the blue range, ii) an underestimation of line fluxes in the far-red range, and iii) an overestimation/underestimation of fluxes for lines bluer/redder to H$\alpha$, respectively,  in the red range. This effect would only affect substantially O$^+$/H$^+$ abundances because {\foii} line fluxes would be overestimated. The effect in N$^+$ would be very small, given the 
proximity of these lines to H$\alpha$. Sulphur ionic abundances are affected by ionization structure effects and can not be discussed on these terms. On the other hand, the effect in the determination of {\Ne} would be negligible because, both {\Ne}({\foii}) and {\Ne}({\fsii}) are computed from line ratios belonging to the same range and very close in wavelength; finally, a small effect would emerge in the determination of {\te}({\fnii}), mainly due to the difference in wavelength between auroral and nebular {\fnii} lines. Taking into account all these possible effects, hereinafter we will not consider aperture 1 computations given the remarkable differences with apertures 2, 3 and 4, and we will consider the weighted average of these three apertures as  representative of the whole nebula.

In last column of Table~\ref{t1}, for comparative purposes, we also present the results for the integrated spectra. The results obtained for the integrated spectra are very similar to those obtained in aperture 1. This is probably due to the effect of apertures 5 to 8 in the collapsed spectrum; these four apertures show higher values of $c$({\hb}) than apertures 2 to 4, hence, mimicking the effect observed in aperture 1.

\section{Quantitative spectroscopic analysis of BD+46\,3474}
\label{section4}
%
We followed a similar strategy as described in \cite{simondiaz10} to perform a detailed, self-consistent 
spectroscopic abundance analysis of BD+463474 by means of the modern stellar atmosphere code 
\fastwind\ \citep{santolayareyetal97, pulsetal05}. In brief, the stellar parameters were derived by comparing the observed H Balmer line profiles and the ratio of \ioni{Si}{iii-iv} line equivalent widths (EWs) with the output from a grid of \fastwind\ models. Then, the same grid of models was used to derive the stellar abundances by means of the curve-of-growth method. 

The projected rotational velocity (\vsini) of the star was derived by means of the {\tt iacob-broad}
procedure implemented in IDL \citep[see notes about its performance in][]{simondiazherrero14} and the measurement 
of the equivalenth width of the metal lines of interest was performed as described in \cite{simondiaz10}. While \cite{simondiaz10} concentrated in the oxygen and silicon abundance determination, in the present study we were able to also extend the abundance analysis to nitrogen, taking advantage of the implementation of a new nitrogen model atom into the \fastwind\ code \citep{riverogonzalezetal11, riverogonzalezetal12}.

In a first step we used the H$_{\gamma}$ line, along with the ratio 
EW(\ioni{Si}{iv}~$\lambda$4116)/EW(\ioni{Si}{iii}~$\lambda$4552) to constrain the effective
temperature and gravity, obtaining 30100$\pm$500, 30500$\pm$500, and 30800$\pm$700~K for 
\grav\,=\,4.1, 4.2, and 4.3 dex, respectively (with \grav\,=\,4.2 dex providing the best
{\em by-eye} fit to H$\gamma$).
An independent determination of these two parameters, together with the helium abundance (Y$_{\rm He}$), the microturbulence (\micro\ ), and the wind-strength Q-parameter (${\dot {\rm M}}$ (R  ${\rm v}_{\infty})^{-3/2}$, Puls et al. 1996) was obtained by performing a HHe spectroscopic analysis using the {\tt iacob-gbat} package \citep{simondiazetal11b}. In this case we used the full set of H and \ioni{He}{i-ii} lines available in the FIES spectrum, obtaining a perfect agreement with the results from the HSi analysis (\Teff\,=\,30100$\pm$1000~K and \grav\,=\,4.2$\pm$0.1 dex), plus Y$_{\rm He}$\,=\,0.11$\pm$0.02, \micro\,$<$\,5 \kms, and log$Q$\,$<$\,-13.5.

%
\begin{table}[t!]
\begin{center}
\caption{\footnotesize First and second columns: Spectroscopic parameters and abundances of Si, O, and N derived through the \fastwind\ analysis of the optical spectrum of BD\,+463474. Third and fourth columns: Some photometric quantities used to compute the physical parameters of the star are also presented for completeness.}
\label{t3}
\begin{tabular}{l c | l c}    
\noalign{\smallskip}
\hline
\hline
\noalign{\smallskip}
\Teff\ (K)                  & \solu{30500}{1000} & V                & 9.74  \\ 
\grav\ (dex)                & \solu{4.2}{0.1}    & (B-V)            & 0.78  \\
Y(He)                       & \solu{0.11}{0.02}  & (B-V)$_{\rm 0}$  & -0.27 \\
log\,Q                      & $<$\,-13.5         & E(B-V)           & 1.05  \\
\micro\ (\kms)	             & $<$\,5	          & $A_{\rm V}$      & 3.25  \\
\vsini\ (\kms)               & $<$15              & $M_{\rm V}$ & -3.0$^{(1)}$  \\
\noalign{\smallskip}
log(Si/H)+12 & \solu{7.51}{0.05}& R (R$_{\odot}$)             & \solu{5.2}{0.2}$^{(1)}$    \\
log(O/H)+12  & \solu{8.73}{0.08} & logL/L$_{\odot}$            & \solu{4.29}{0.03}$^{(1)}$  \\
log(N/H)+12  & \solu{7.86}{0.05} & M$_{\rm sp}$ (M$_{\odot}$)  & \solu{15}{2}$^{(1)}$       \\
\hline
\hline
\end{tabular}
\begin{description}
\item[] $^{(1)}$ Corresponding to a distance of 800 pc (see notes in Appendix A). 
\end{description}
\end{center}
\end{table}

\begin{figure}[!t]
\centering
\includegraphics[width=8.7cm,angle=0]{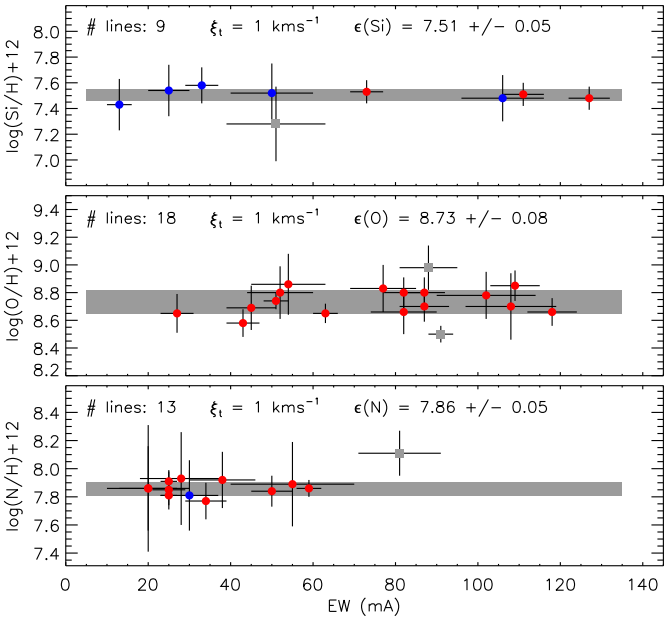}
\caption{ Final log(X/H) vs. EW diagrams for Si, O, and N, resulting from the abundance
analysis of BD\,+46~4374. The microturbulence providing a minimum dispersion of line-by-line
abundances is indicated at the top of each panel along with the associated mean and abundance 
dispersion. Red and blue dots indicate lines from different ions (red dots: \ion{Si}{iii}, \ion{O}{ii} and \ion{N}{ii}; blue dots: \ion{Si}{iv} and \ion{N}{iii}); grey dots correspond to those lines excluded from the analysis (see also Sect. \ref{section4} 
and Table \ref{t4}). The grey horizontal band correspond to the adopted value and its uncertainty. }
\label{f4}
\end{figure}

The final set of stellar parameters derived spectroscopically (along with the corresponding uncertainties) is summarized in Table~\ref{t3}. For completeness, we also include in Table~\ref{t3} 
 the derived Si, O and N abundances (see below) as well as other stellar
parameters of interest for further studies of the Cocoon nebula and its ionizing source such as the 
radius, luminosity and spectroscopic mass (see notes about how these parameters were derived in Appendix A). 
The spectroscopic parameters were then fixed for the subsequent Si, O and N abundance analysis. 
We present in Table~\ref{t4} the considered set of \ioni{Si}{iii-iv}, \ioni{O}{ii}, and \ioni{N}{ii-iii}
diagnostic lines, along with the measured equivalent widths and corresponding line-by-line 
abundances\footnote{$\epsilon_{\rm X}$\,=\,log(X/H)+12}
(plus the associated uncertainties). The same information is represented graphically in Figure~\ref{f4}.

In the three cases, a very low value of microturbulence (\micro\,=\,1$\pm$1~\kms) is required to obtain a zero slope in the $\epsilon_{\rm X}$\,--\,EW diagrams. We have marked in bold, and as grey squares in Figure~\ref{f4}, those lines whose abundances deviates more than 2$\sigma$ from the resulting distribution of abundances. These lines were excluded from the final computation of mean abundances -- $\epsilon_{\rm X}$ -- and standard deviations -- $\Delta\epsilon_{\rm X}$($\sigma$) --. We also indicate in Table~\ref{t4} the uncertainty associated with a change of $\pm$\,1~\kms\ in microturbulence -- $\Delta\epsilon_{\rm X}$(\micro) -- and, for the case of oxygen, the effect of modifying \Teff\ and \grav\ in $\pm$~1000 K and 0.1~dex, respectively -- $\Delta\epsilon_{\rm X}$(SP) 
-- .

%
\begin{table}[!h]
{\scriptsize
\begin{center}
\caption{\small Results from the abundance analysis of
BD\,+463474 (B0.5\,V). Values in bold indicate lines whose abundances deviates more 
than 2$\sigma$ from the resulting distribution of abundances.} 
 
\label{t4}
\begin{tabular}{ccccc}
\noalign{\smallskip}
\hline\hline
\noalign{\smallskip}
BD\,+463474 &
\multicolumn{3}{c}{\Teff\,=\,30500 K, \grav\,=\,4.2 dex} & \micro(Si)=1  \\
\hline
\noalign{\smallskip}
Line & $EW$ & $\Delta$$EW$ & $\epsilon_{\rm Si}$ & $\Delta\epsilon_{\rm Si}$ \\
 & (m\AA) & (m\AA) & (dex) & (dex) \\
\hline
\noalign{\smallskip}
\ioni{Si}{iii}$\lambda$4552   &   127     &      5    &       7.48 &         0.09 \\
\ioni{Si}{iii}$\lambda$4567   &   111     &      5    &       7.51 &         0.09 \\
\ioni{Si}{iii}$\lambda$4574   &    73     &      4    &       7.53 &         0.09 \\
 \ioni{Si}{iv}$\lambda$4116   &   106     &     10    &       7.48 &         0.18 \\
 \ioni{Si}{iv}$\lambda$4212   &    33     &      4    &       7.58 &         0.14 \\
 \ioni{Si}{iv}$\lambda$4631   &    50     &     10    &       7.52 &         0.23 \\
{\bf \ioni{Si}{iv}$\lambda$4654}   &    51      &    12     &      7.28  &        0.29 \\
 \ioni{Si}{iv}$\lambda$6667   &    13     &      3    &       7.43 &         0.20 \\
 \ioni{Si}{iv}$\lambda$6701   &    25     &      5    &       7.54 &         0.20 \\
\hline
\noalign{\smallskip}
\multicolumn{3}{c}{} &
 {\bf $\epsilon_{\rm Si}$\,=\, 7.51} &
 \multicolumn{1}{r}{$\Delta\epsilon_{\rm Si}$($\sigma$)\,=\,0.05} \\
 &  \multicolumn{2}{r}{$\Delta$\micro(Si)\,=\,1}  & $\Rightarrow$ &
 \multicolumn{1}{r}{$\Delta\epsilon_{\rm Si}$(\micro)\,=\,0.06} \\
\end{tabular}
\begin{tabular}{ccccc}
\noalign{\smallskip}
\hline\hline
\noalign{\smallskip}
BD\,+463474 &
\multicolumn{3}{c}{\Teff\,=\,30500 K, \grav\,=\,4.2 dex} & \micro(O)=1 \\
\hline
\noalign{\smallskip}
\ioni{O}{ii}$\lambda$3945   &    54    &       9     &      8.86 &       0.22 \\
\ioni{O}{ii}$\lambda$3954   &    77    &       8     &      8.83 &       0.17 \\
\ioni{O}{ii}$\lambda$4317   &   102    &      12     &      8.78 &       0.17 \\
\ioni{O}{ii}$\lambda$4319   &    87    &       6     &      8.70 &       0.11 \\
\ioni{O}{ii}$\lambda$4366   &    82    &       8     &      8.66 &       0.16 \\
\ioni{O}{ii}$\lambda$4414   &   118    &       6     &      8.66 &       0.10 \\
\ioni{O}{ii}$\lambda$4416   &   109    &       6     &      8.85 &       0.11 \\
\ioni{O}{ii}$\lambda$4452   &    52    &       8     &      8.80 &       0.19 \\
{\bf \ioni{O}{ii}$\lambda$4638}   &    88    &       7     &      8.98 &       0.16 \\
{\bf \ioni{O}{ii}$\lambda$4641}   &    91    &       3     &      8.50 &       0.06 \\
\ioni{O}{ii}$\lambda$4661   &    87    &       5     &      8.80 &       0.11 \\
\ioni{O}{ii}$\lambda$4676   &    82    &       5     &      8.80 &       0.11 \\
\ioni{O}{ii}$\lambda$6721   &    51    &       3     &      8.74 &       0.06 \\
\ioni{O}{ii}$\lambda$4076   &   108    &      11     &      8.70 &       0.24 \\
\ioni{O}{ii}$\lambda$4891   &    27    &       4     &      8.65 &       0.14 \\
\ioni{O}{ii}$\lambda$4906   &    45    &       6     &      8.69 &       0.16 \\
\ioni{O}{ii}$\lambda$4941   &    43    &       4     &      8.58 &       0.10 \\
\ioni{O}{ii}$\lambda$4943   &    63    &       3     &      8.65 &       0.07 \\
\hline
\noalign{\smallskip}
 \multicolumn{3}{c}{} &
 {\bf $\epsilon_{\rm O}$\,=\, 8.73} &
 \multicolumn{1}{r}{$\Delta\epsilon_{\rm O}$($\sigma$)\,=\,0.08} \\
 &  \multicolumn{2}{r}{$\Delta$\micro(O)\,=\,1}  & $\Rightarrow$ &
 \multicolumn{1}{r}{$\Delta\epsilon_{\rm O}$(\micro)\,=\,0.03} \\
 \multicolumn{3}{r}{$\Delta$\Teff\,=\,1000, $\Delta$\grav\,=\,0.1} & $\Rightarrow$ &
 \multicolumn{1}{r}{$\Delta\epsilon_{\rm O}$(SP)\,=\,0.08} \\
\end{tabular}
\begin{tabular}{ccccc}
\noalign{\smallskip}
\hline\hline
\noalign{\smallskip}
BD\,+463474 &
\multicolumn{3}{c}{\Teff\,=\,30500 K, \grav\,=\,4.2 dex} & \micro(N)=1 \\
\hline
\noalign{\smallskip}
{\bf \ioni{N}{ii}$\lambda$3995}  &     81    &      10    &       8.11    &      0.16 \\
  \ioni{N}{ii}$\lambda$4607  &     20    &      10    &       7.86    &      0.45 \\
  \ioni{N}{ii}$\lambda$4621  &     20    &       7    &       7.86    &      0.30 \\
  \ioni{N}{ii}$\lambda$4630  &     55    &      15    &       7.89    &      0.30 \\
  \ioni{N}{ii}$\lambda$4643  &     25    &       2    &       7.81    &      0.07 \\
  \ioni{N}{ii}$\lambda$4447  &     38    &       8    &       7.92    &      0.20 \\
  \ioni{N}{ii}$\lambda$5045  &     28    &      10    &       7.93    &      0.33 \\
  \ioni{N}{ii}$\lambda$5666  &     34    &       5    &       7.77    &      0.13 \\
  \ioni{N}{ii}$\lambda$5676  &     25    &       2    &       7.91    &      0.07 \\
  \ioni{N}{ii}$\lambda$5679  &     59    &       3    &       7.86    &      0.06 \\
  \ioni{N}{ii}$\lambda$5007  &     25    &       4    &       7.85    &      0.14 \\
  \ioni{N}{ii}$\lambda$5005  &     50    &       5    &       7.84    &      0.11 \\
 \ioni{N}{iii}$\lambda$4634  &     30    &       7    &       7.81    &      0.25 \\
\hline
\noalign{\smallskip}
 \multicolumn{3}{c}{} &
 {\bf $\epsilon_{\rm N}$\,=\, 7.86} &
 \multicolumn{1}{r}{$\Delta\epsilon_{\rm N}$($\sigma$)\,=\,0.05} \\
 &  \multicolumn{2}{r}{$\Delta$\micro(N)\,=\,1}  & $\Rightarrow$ &
 \multicolumn{1}{r}{$\Delta\epsilon_{\rm N}$(\micro)\,=\,0.02} \\
\end{tabular}
\end{center}
}
\end{table}
%

%
\section{Discussion}
\label{section5}
%


\subsection{Nebular and stellar abundances: Cocoon vs. Orion and M\,43}
\label{section51}

\begin{table*}
\begin{center}
\caption{Comparison of derived stellar and gas-phase nebular abundances obtained for the Cocoon nebula, M\,43 and the Orion nebula. All the CEL abundances are for $t^2=0$. Abundances from other B-type stars in the Solar neighborhood are also quoted for reference.}
\label{t5}
\begin{tabular}{lccccccc}
\noalign{\smallskip} \noalign{\smallskip} \noalign{\hrule} \noalign{\smallskip}
	&  \multicolumn{2}{c}{Cocoon$^{\rm a}$} & \multicolumn{2}{c}{Orion} & M\,43 & Solar neighborhood	\\
\cline{2-3} \cline{4-5} \cline{6-7}
\noalign{\vskip3pt} 
       & Nebular (CEL) & B stars & Nebular (CEL)$^{\rm b}$ & B stars &Nebular (CEL)$^{\rm f}$ & B stars$^{\rm g}$ \\
\noalign{\hrule} \noalign{\smallskip}
Si      & $-$          		& 7.51$\pm$0.05 	& $-$ 			& 7.51$^{\rm c}$/7.50$^{\rm d}$$\pm$0.05 		& $-$ 		& 7.50$\pm$0.05\\
O	& 8.52$\pm$0.05	& 8.73$\pm$0.08	& 8.54$\pm$0.03	& 8.73$^{\rm c}$/8.77$^{\rm d}$$\pm$0.04 		& 8.53$\pm$0.09& 8.76$\pm$0.05\\
N	& 7.81$\pm$0.03	& 7.86$\pm$0.05	& 7.73$\pm$0.09	& 7.82$^{\rm d}$$\pm$0.07 				& 7.82$\pm$0.04& 7.79$\pm$0.04\\
S	& 6.81$\pm$0.04	& $-$			& 6.83$\pm$0.04	& 7.15$^{\rm e}$$\pm$0.05 				& 6.95$\pm$0.09&  $-$\\
\noalign{\smallskip} \noalign{\hrule} \noalign{\smallskip}
\end{tabular}
\begin{description}
\scriptsize
{\bf Notes:} $^{(a)}$ This work; $^{(b)}$ Reanalysis of the Orion nebula spectrum by \cite{estebanetal04} using the same atomic dataset as in this work (see Table~\ref{t2}); $^{(c)}$ \citet{simondiaz10}, similar analysis as in this work using \fastwind; $^{(d)}$ \citet{nievasimondiaz11}, reanalysis of the spectra from \cite{simondiaz10} using similar techniques as in \cite{nievaprzybilla12}; $^{(e)}$\citet{daflonetal09}; $^{(f)}$ Reanalysis of M\,43 spectrum by \cite{simondiazetal11} using the same atomic dataset as in this work; $^{(g)}$ \cite{nievaprzybilla12}.
\end{description}
\end{center}
\end{table*}

The study of interstellar absorption lines of the cold gas by \citet{sofiameyer01} 
demonstrated that the local ISM out to 1.5~Kpc from the Sun is chemically homogeneous to the 10\% level.
A similar result has been more recently obtained by \cite{nievaprzybilla12} from a very thoughtful abundance
analysis of a sample of 29 early B-type stars located 500~pc around the Sun. Given its distance to the
Sun ($\sim$1~Kpc), the Cocoon nebula and BD+46~3474 are expected to share the same chemical composition 
as the Orion nebula (d$\sim$450 pc) and other B-type stars in the Solar vicinity. For a discussion about the comparion of chemical abundances from B-type stars in the solar neighbourhood and abundances obtained in the Sun, we refer the reader to the study of \citet{nievaprzybilla12} who made this comparison in the framework of recent observational data and Galactic chemical and kinematical evolution models.

In the first two columns of Table~\ref{t5} we present a summary of the Cocoon gas-phase nebular and stellar abundances that will be considered hereafter. As discussed in Sect. \ref{section34}, we will assume the weighted average of the gas-phase abundances for apertures 2 to 4 as representative of the nebula. For comparative purposes, we also quote: 
\begin{itemize}
 \item the N, O and S abundances resulting from a reanalysis of the Orion nebula and M43 spectra presented in \cite{estebanetal04} and \cite{simondiazetal11b} using the same atomic datset as in this paper;
 \item the Si, O, and N abundances derived by \cite{simondiaz10}, \cite{nievasimondiaz11} for B-type stars in the Orion star forming region and by \cite{nievaprzybilla12} for B-type stars in the Solar vicinity; and
 \item the S abundances derived by \cite{daflonetal09} for B-type stars in the Orion star forming region.
 \end{itemize}

The agreement between the Si, O and N abundances derived from the spectroscopic analysis of BD+46~3474 and the recent determinations of abundances in B-type stars in the Orion OB1 association and, more generally, the solar neighborhood is quite remarkable.

The comparison of gas-phase abundances in the Cocoon nebula and the Orion star-forming region (Orion nebula+M\,43) is also almost perfect when the same atomic datasets are used (see, however, the effect of assuming different available
atomic data for sulphur in Sect.~\ref{section52}). Particularly, the comparison with the results obtained for M\,43 is of special interest, owing to no ICFs are needed for computing O, N and S abundaces in M\,43 \citep[see][]{simondiazetal11}. Using the same atomic dataset than in this work, the abundances of O and N are in excellent agreement between both nebulae (see Table~\ref{t5}). The differences between S abundances in the Cocoon nebula and in M\,43 cannot be attributed to atomic data, nor to the use of an ICF; however, we have used different lines for computing S$^{2+}$/H$^+$ ratio: in the Cocoon nebula we used the bright nebular lines at $\lambda\lambda$9069,9531 while in M\,43\footnote{The spectra of M43 used by \citet{simondiazetal11} do not cover the near infrared zone of the spectrum where the bright nebular [SIII] lines lie.} we used the faint and extremely temperature dependent auroral line at $\lambda$6312. As we have assumed the same temperature for S$^+$ and S$^{2+}$ regions (\Te([\ion{N}{ii}]) for 
the 
Cocoon nebula and \Te([\ion{O}{ii}]) for M\,43), differences in the true temperature in the zones where the different ions are present may  explain the observed discrepancy. This fact remarks the importance of using consistent sets of lines, physical conditions, atomic data and ICFs when comparing abundances obtained for the same element in different objects.

In the case of the Orion nebula, there is a small difference between N abundances, which may be perfectly
explained due to uncertainty in the ICF(N) which has to be assumed for the case of the Orion nebula. 
Indeed, we only indicate in Table~\ref{t5} the value provided by \citet{garciarojasesteban07}; however, \cite{estebanetal04} and \citet{simondiazstasinska11} proposed another two possible values of the total gas-phase N abundance in the Orion nebula (7.65$\pm$0.09 and 7.92$\pm$0.09, respectively). 
The three values result from the analysis of the same spectrum but a different assumption of the ICF(N). 
In this context, we highlight the importance of the Cocoon nebula and M\,43 for the comparison of nitrogen 
abundances derived from the analysis of the nebular and stellar spectra (Sect.~\ref{section52}), since 
in these cases no ICF(N) is needed to obtain the total nebular nitrogen abundance. The results obtained for these two nebulae strongly favour the ICF(N) used by \citet{garciarojasesteban07} for computing total N abundance in the Orion nebula.

\subsection{Nebular $vs.$ stellar abundances in the Cocoon nebula}
\label{section52}

Before starting the comparison of nebular and stellar abundances in the Cocoon nebula, we want to
briefly summarize the main results of a similar study performed by \citet{simondiazstasinska11} in the
Orion star forming region. They used the chemical abundance study of the Orion star forming region from B-type stars \citep{simondiaz10, nievasimondiaz11} to compare the derived abundances with those obtained for non-refractory elements (C, N, O and Ne) in the most detailed study of the gas-phase chemical abundances on the Orion nebula \citep{estebanetal04}. The main conclusion of these authors is that oxygen abundance derived from CELs (corrected from depletion onto dust grains) in the Orion nebula is irreconcilable with that derived from B-type stars. On the other hand, they find that N and Ne gas phase abundances and C gas phase$+$dust abundances from CELs seemed to be consistent with those derived in B-type stars \citep[see Fig.~1 of][]{simondiazstasinska11}. In addition, these authors find that oxygen gas phase+dust abundances derived from optical recombination lines (ORLs) agree very well with oxygen abundances derived in the stars. 

In the study of the Cocoon nebula presented here we concentrate on the comparison of nebular and stellar abundances for O, N and S. As indicated in Sects. \ref{section34} and \ref{section51}, one important point of this study (compared to the case of the Orion nebula) is that the total abundances of the three investigated elements are obtained without the necessity of any ICF. Although we only have access to nebular abundances derived from CELs (no ORLs are detected), we will also include in our discussion how CEL abundances corrected from the presence of possible temperature fluctuations compare to the stellar ones. In particular, since we cannot directly compute the $t^2$ parameter from our observations, we will assume two cases: a canonical value of $t^2$=0.035 \citep[which is an average value in Galactic {\hii} regions, see][]{garciarojasesteban07}, and the value derived for the Orion
nebula \citep[$t^2$=0.022, ][]{estebanetal04}.

A meaningful comparison of nebular and stellar abundances first requires the nebular gas-phase abundances (Tables~\ref{t1} and \ref{t5}) to be corrected from possible depletion onto dust grains.
Several authors have estimated the oxygen deplection factor
in the Orion nebula by comparing abundances of the refractory elements Mg, Si and Fe in the gas phase with those found in the atmospheres of B stars of the Orion cluster \citep[e.g.][]{estebanetal98, mesadelgadoetal09b, simondiazstasinska11}; these authors found oxygen depletions between 0.08 and 0.12 dex. For a detailed discussion on the computations of such depletions, we refer the reader to \citet{simondiazstasinska11}.
Lacking for the whole bunch of information needed to perform a similar computation in the Cocoon nebula, we decided to
adopt a canonical value of 0.10 dex as representative of the oxygen depletion in this nebula and consider an associated uncertainty of $\pm$0.02 dex.
Nitrogen is expected not to be a major constituent of dust in \hii\ regions \citep{jenkins14}; therefore, no correction is needed.
For sulphur, the situation is more complicated; although for a long time sulphur was thought not to be depleted 
onto dust grains \citep[see e.g.][]{sofiaetal94}, recently, some authors \citep{jenkins09, whitesofia11} drew attention about the risks of assume sulphur as a standard for what should be virtually zero depletion, especially for some sight lines. Unfortunately, there is a lack of quantitative studies on the sulphur depletion onto dust grains, that makes this an open question that needs to be addressed in the future by using high-quality interstellar abundance measurements. We hence assume no dust correction for sulphur, but keep in mind the abovementioned argument.

\subsubsection{Oxygen}

Similarly to the case of the Orion nebula, the derived gas+dust oxygen abundance resulting from CELs and a $t^2$=0 (8.62 $\pm$0.05) is remarkably different to the O abundance obtained from the spectroscopic analysis
of the central star.
If we consider as valid the assumption that temperature fluctuations are affecting the determination of ionic chemical abundances using CELs \citep{peimbert67, peimbertcostero69}, and the canonical value of $t^2$=0.035 
we would obtain that total gas+dust nebular abundance would reach 12+log(O/H)=8.86, which is now much larger than that obtained from stars. While this result could be used as an argument against the temperature fluctuation scheme,
we must remind that we considered a value of $t^2$ that may not be representative of the actual value in the Cocoon nebula.
In particular, if the value of $t^2$ derived for the Orion nebula is considered (0.022), the resulting gas+dust oxygen abundance would be 12+log(O/H)=8.75, in much better agreement with the stellar one.


\subsubsection{Nitrogen}

As \citet{simondiazstasinska11} argued, if the RL-CEL abundance discrepancy were caused by temperature fluctuations, as suggested by \citet{peimbertetal93}, one should observe the same kind of bias in the CEL abundances of the other elements. 
They do not find other elements such as N, C and Ne following the same behaviour as oxygen;
however, they also claim that the derived total gas-phase abundances of C, N and Ne in the Orion nebula are much less accurate. This is mainly due to the uncertainties on the adopted ICFs. In the Cocoon nebula, no ICF correction is needed to be applied to compute the total N gas-phase abundance. 
This is due to the low excitation of the nebula, that prevents the ionization of N$^+$ to N$^{2+}$. From the comparison of the total nebular N abundance obtained from CELs (and $t^2$=0), 12+log(N/H)=7.81$\pm$0.03 and that obtained from the analysis of the central star, 12+log(N/H)=7.86$\pm$0.05, we can conclude that the stellar one is slightly higher but both values are consistent within the uncertainties. 
In this case, the CEL+$t^2$ abundances that result from assuming a $t^2$=0.035 (canonical) or 
0.022 (Orion nebula) are 8.02 and 7.93, respectively\footnote{Note that the correction to the abundances derived from {\fnii} CELs is much lower than that derived from {\foii} CELs. In particular, this correction depends on the wavelength of the used lines for abundance calculations, being larger for bluer lines, such as {\foii} $ \lambda\lambda$3726+29 and lower for redder lines, such as {\fnii} $\lambda\lambda$6548+83.}. The later option 
is also in agreement with the stellar solution. 

For completeness in this section we must write a word of caution regarding the stellar nitrogen abundance. 
Spectroscopic analysis of early-B type main sequence stars in the last years have shown an increasing 
observational evidence of the existence of a non-negligible percentage of narrow lined (low \vsini, but 
not neccesarily fast rotators seen pole-on) targets among these stars showing nitrogen enhancement in their photospheres \cite[e.g.,][]{morel06, hunter08}. This result
warns us about the danger of extracting any conclusion from the direct comparison of nebular an stellar
abundances based in one target. We hence must consider the derived nitrogen abundance in BD+46~3474 as an 
upper limit to be compared with the nebular abundance, specially in view of the nitrogen
abundance obtained for this star in comparison with other stars in the solar neighborhood (see Table~\ref{t5}).

\subsubsection{Sulphur}

Given the low excitation of the Cocoon nebula, that prevents the presence of ionization species of S
higher than S$^{\rm 2+}$, we skip the uncertainty associated with the use of an ICF to compute the total nebular sulphur abundance.\footnote{Taking into account the lack of O$^{2+}$ in the nebula, and the similarity between ionization potentials of O$^{+}$ (35.12 eV) and S$^{2+}$ (34.83 eV), this seem to be a reasonable conclusion.} However, there are still a couple of issues that makes the comparison of nebular and stellar abundances for this element still uncertain. First, the remarkable difference in the computed abundances when assuming 
different atomic datasets. To illustrate this we have recomputed the Coocon nebular sulphur abundance using the same atomic data for this element as \cite{garciarojasesteban07}\footnote{The atomic dataset used by these authors was the following: Collision strengths by \citet{ramsbottometal96} for S$^+$ and \citet{tayalgupta99} for S$^{2+}$. Transition Probabilities by \citet{keenanetal93} for S$^+$ and \citet{mendozazeippen82b} and \citet{kaufmansugar86} for S$^{2+}$}. Although it was not commented in Section~\ref{section51}, the value proposed by these authors for the sulphur abundance in the Orion nebula was 7.04$\pm$0.04, a factor 1.6 larger than the abundance indicated in Table~\ref{t5}. In the case of the Cocoon nebula, the abundance is modified from 6.81 to 6.90 (i.e. 20\% difference). 
Second,
as discussed above, there are doubts about 
the depletion of sulphur on dust \citep{jenkins09} and some amount of it may be in the form of dust grains.
In addition, we still do not have implemented and tested a sulphur model atom to be used with \fastwind\, and could not derive the S abundance associated with BD+46\,3474; however, given the good match between the Si, O and N abundances in BD+46~3474 and other stars B-type in the Orion star forming region (Sect. \ref{section51}) we consider as a still valid exercise the comparison of our nebular sulphur abundance with those obtained by \cite{daflonetal09}\footnote{A more
recent paper by \cite{irrgangetal14} include three of the Orion stars analyzed in \cite{simondiaz10} using
a similar technique and atomic data as \cite{nievasimondiaz11}. The resulting sulphur abundances are in
good agreement with the values obtained by \cite{daflonetal09}. }.

From the comparison of the sulphur abundances presented in Table~\ref{t5} we can conclude 
that there is a clear discrepancy (by more than 0.3 dex) between the nebular CEL ($t^2$=0) and stellar abundances.
If we consider the presence of temperature fluctuations, and assume the canonical value for Galactic 
{\hii} regions and that of the Orion nebula, we would overcome a large part of the discrepancy, reaching to values of 12+log(S/H)=7.07 and 6.96, respectively (to be compared with 7.15$\pm$0.05). While these values are in much better agreement with 
the stellar abundance, the $t^2$=0.022 solution (the one for which we find better agreement in the case of O and N) is still far away from the value resulting from the analysis of the stellar spectra.

Interestingly, a perfect agreement between nebular and stellar sulphur abundances would be obtained if the atomic dataset considered by \cite{garciarojasesteban07} is assumed and combined with a $t^2$=0.035. However, while still a valid option, this possibility is highly
speculative and far from being considered as a valid scientific argument supporting any conclusion.

%
\begin{figure}[!ht]
\centering
\includegraphics[width=5.6cm,angle=90]{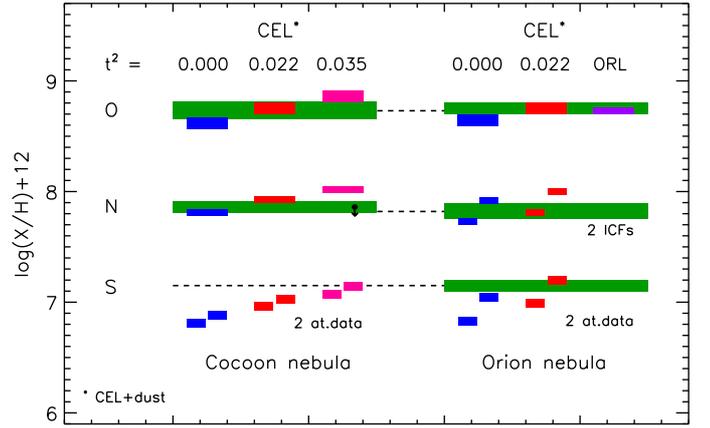}
\caption{Comparison of O, N, and S abundances obtained from B-type stars (green) and nebular (gas+dust) abundances derived from CEL and different values of $t^2$ (blue, red, and magenta). Violet box indicates O abundance obtained from ORLs in the Orion nebula by \citet{estebanetal04}. 
The height of the boxes represents the uncertainties. Left and right columns corresponds to the Cocoon nebula and the Orion nebula, respectively. For sulphur, the resulting abundances using two different atomic dataset are indicated; for nitrogen in the Orion nebula, the abundances computed by assuming two different ICFs are shown. See text for more details.}
\label{f5}
\end{figure}

\begin{table*}[t!]
\begin{center}
\caption{\footnotesize Summary of the abundances presented in Figure~\ref{f5} .  }
\label{t6}
\scriptsize
\begin{tabular}{c c c c c c c c c c}    
\noalign{\smallskip} \noalign{\smallskip} \noalign{\hrule} \noalign{\smallskip}
	&  \multicolumn{4}{c}{Cocoon} & & \multicolumn{4}{c}{Orion} \\
\cline{2-5} \cline{7-10}
\noalign{\vskip3pt} 
       & \multicolumn{3}{c}{Nebular gas+dust (CEL)} & B stars & & \multicolumn{2}{c}{Nebular gas+dust (CEL)} & Nebular  & B stars  \\
 & & & & & & & &gas+dust (ORL) & \\
\cline{2-4} \cline{7-9}
\noalign{\vskip3pt} 
       & $t^2$=0.00 & $t^2$=0.022 & $t^2$=0.035 & & & $t^2$=0.00 & $t^2$=0.022  &  & \\
\noalign{\hrule} \noalign{\smallskip}
O	& 8.62$\pm$0.05 & 8.75$\pm$0.05 & 8.86$\pm$0.05	& 8.73$\pm$0.08	& & 8.64$\pm$0.03	& 8.75$\pm$0.03 & 8.73$\pm$0.03 & 8.75$\pm$0.05\\
N	& 7.81$\pm$0.03 & 7.93$\pm$0.03 & 8.02$\pm$0.03	& 7.86$\pm$0.05	& & 7.73/7.92$\pm$0.03$^{1}$	& 7.81/8.00$\pm$0.03$^{1}$ & $-$ & 7.82$\pm$0.07 \\
S	& 6.81/6.88$\pm$0.04$^{2}$ & 6.96/7.03$\pm$0.04$^{2}$   & 7.07/7.14$\pm$0.04$^{2}$  & $-$	&		
& 6.83/7.04$\pm$0.04$^{2}$	& 6.99/7.20$\pm$0.04$^{2}$ & $-$ & 7.15$\pm$0.05\\
\noalign{\smallskip} \noalign{\hrule} \noalign{\smallskip}
\end{tabular}
\begin{description}
\scriptsize
\item[] $^{1}$ Two ICFs used (see text). 
\item[] $^{2}$ Two sets of atomic data used (see text). 
\end{description}
\end{center}
\end{table*}

\subsubsection{Final remarks}

In Fig.~\ref{f5} we illustrate all the discussion above showing the comparison between the abundances of O, N and S in the Cocoon nebula obtained using CELs and $t^2=0$ (blue boxes), CELs and $t^2=0.022$ (the Orion nebula value, red boxes) and CELs and $t^2=0.035$ (canonical $t^2$ value for Galactic {\hii} regions, magenta) with the abundances in the central star BD+46\,3474 (green boxes). For comparison we also represent the values for CELs and $t^2=0$ and $t^2=0.022$ as well as oxygen ORLs values (cyan box) for the Orion nebula. 
The height of the boxes represents the adopted uncertainties. All these numbers are summarized in Table~\ref{t6}.  
As a general result, for O and N, it is clear that abundances considering small temperature fluctuations ($t^2=0.022$, similar to what found in the Orion nebula) agree much better with what obtained from stars than those considering pure CELs with $t^2=0$.  On the other hand, typical values of $t^2=0.035$ found in Galactic {\hii} regions seem to overestimate nebular O and N abundance. As representative of the ICF problem, in Fig.~\ref{f5} we show two values of the Orion N abundance, assuming the ICF by \citet{garciarojasesteban07} (left) and \citet{simondiazstasinska11} (right); it is evident that by using different ICFs one can reach radically different conclusions. 

In the case of sulphur, the situation is puzzling. Although some increase of the S abundance owing to depletion onto dust can not be ruled out, it seems that the results for CELs and $t^2=0$ are far from stellar abundances. Additionally, the abundances are highly dependent on the selected atomic dataset (especially for the Orion nebula where S$^{2+}$ is dominant). We have computed sulphur nebular abundances using atomic data from Table~\ref{t2} (left) and atomic data used in \citet{garciarojasesteban07}. In general, by using the atomic dataset of Table~\ref{t2}, sulphur nebular abundances are going to be far from the stellar results assuming $t^2=0.022$ in both, the Cocoon and the Orion nebula, but can be in agreement assuming $t^2=0.035$, which has been considered too high from the analysis of O and N data. On the other hand, assuming the atomic dataset of \citet{garciarojasesteban07} the situation changes and then one can reconcile sulphur nebular and stellar abundances by assuming $t^2=0.022$. Of course,
 this result do not necessarily favours one given dataset, but warns about the influence of atomic data in our results \citep[see][for a critical review on the use of nebular atomic data]{luridianaetal11, luridianagarciarojas12}.

All the results above do not necessary imply the existence of such temperature fluctuations, whatever the physical origin of such fluctuations, but they warn us about the use of pure CELs as a proxy for computing chemical abundances in photoionized regions. Additionally special care should be taken into account when selecting an ICF scheme and/or an atomic dataset, owing to a bad choice could reach to large uncertainties on the nebular chemical abundances and hence, to incorrect interpretations.

We have to emphasize the importance of including more elements to compare in future studies (mainly C, Ne and Ar). The main problem with these elements is that we would always need an ICF to compute the total nebular abundance from optical spectra. This problem may be circumvent using multiwavelength studies including UV and IR lines, but this is very difficult for very extended objects, such as Galactic {\hii} regions owing to the different apertures used for UV, optical and IR spectrographs, which may introduce non-negligible ionization structure effects. A detailed set of photoionization models covering as much as possible the {\hii} regions parameter space is needed to build a consistent set of ICFs, as it has been recently done for planetary nebulae by \citet{delgadoingladaetal14}.

%
\section{Summary and conclusions}
\label{section6}
%

The Cocoon Nebula (IC\,5146) $-$a roundish {\hii} region ionized by a single B0.5\,V star (BD+46\,3474)$-$ seems to be an ideal object to compare stellar and nebular chemical abundances and then check the abundance determinations methods in the field of {\hii} regions and massive stars. 

We collect a set of high quality observations comprising the optical spectrum of BD+46\,3474 (the main ionizing source), along with long-slit spatially resolved nebular spectroscopy of the nebula .

In this paper we present the nebular abundance analysis of the spectra extracted from apertures located at various distances from the central star in the Cocoon nebula, as well as a quantitative spectroscopical analysis of the ionizing central star BD+46\,3474.

We performed a detailed nebular emipirical analysis of 8 apertures extracted from 
a long-slit located to the north-west of BD+46\,3474. We obtained the spatial distribution of the physical conditions (temperature and density) and ionic abundances of \ion{O}{$^+$}, \ion{N}{$^+$}, \ion{S}{$^+$} and \ion{S}{$^{2+}$}. Owing to the extremely low ionization degree of the Cocoon nebula, we can determine total abundances directly
from observable ions, eliminating the uncertainties resulting of assuming an ICF scheme, which are especially significant for the case of N. In particular, the N abundance is in  complete agreement with that  determined by \citet{simondiazetal11} for M\,43, a local \ion{H}{ii} region with a similar low ionization degree. 

By means of a quantitative spectroscopic analysis of the optical spectrum of BD+46\,3474
with the stellar atmosphere code FASTWIND we derived for this B0.5\,V star 
\Teff\,=\,30500$\pm$1000 K and \grav\,=\,4.2$\pm$0.1. and chemical abundances of Si, O and N (in 12+log(X/H)) of 7.51$\pm$0.05, 8.73$\pm$0.08 and 7.86$\pm$0.05, respectively.

From the comparison of O, N and S abundances in the nebula and in its central star we conclude that: i) abundances derived from CELs are, in general, lower to those found in stars for the same element; ii) considering moderate temperature fluctuations, similar to what found in the Orion nebula ($t^2=0.022$), and dust depletion for O, we would reconcile the abundances in the nebula and the central star for O and N. For S, the results are somewhat puzzling and points to 
different conclusions depending on the atomic dataset adopted for computing the ionic abundances. 

As a future step, this type of study should be extended to other elements and {\hii} regions with the aim of looking for systematic effects in the nebular/stellar abundances. Multiwavelength nebular studies taking into account aperture effects, and/or a new set of theoretical ICFs from a complete grid of {\hii} region photoionization models, as well as multielement abundance studies from a large number of massive stars in the same star forming region would minimize uncertainties and probably would shed some light on this still poorly explored topic.

\begin{acknowledgements}
%
This work received financial support from the Spanish Ministerio  de Educaci\'on y Ciencia (MEC) under project AYA2011-22614. JGR and SSD acknowledge support from Severo Ochoa excellence program (SEV-2011-0187) postdoctoral fellowships. We thank the anonymous referee for his/her suggestions.
%
\end{acknowledgements}
%


\begin{thebibliography}{61}
\expandafter\ifx\csname natexlab\endcsname\relax\def\natexlab#1{#1}\fi

\bibitem[Aller et al.(1982)]{aller82} 
Aller, L.~H., Appenzeller, I., Baschek, B., et al.\ 1982, Landolt-Bornstein: Group 6: 
Astronomy,  

\bibitem[{{Bresolin} {et~al.}(2009){Bresolin}, {Gieren}, {Kudritzki},
  {Pietrzy{\'n}ski}, {Urbaneja}, \& {Carraro}}]{bresolinetal09}
{Bresolin}, F., {Gieren}, W., {Kudritzki}, R.-P., {et~al.} 2009, ApJ, 700, 309

\bibitem[Brott et al.(2011)]{brott11} 
Brott, I., de Mink, S.~E., Cantiello, M., et al.\ 2011, \aap, 530, A115 

\bibitem[{{Carigi} {et~al.}(2005){Carigi}, {Peimbert}, {Esteban}, \&
  {Garc{\'{\i}}a-Rojas}}]{carigietal05}
{Carigi}, L., {Peimbert}, M., {Esteban}, C., \& {Garc{\'{\i}}a-Rojas}, J. 2005,
  ApJ, 623, 213

\bibitem[{{Chiappini} {et~al.}(2003){Chiappini}, {Romano}, \&
  {Matteucci}}]{chiappinietal03}
{Chiappini}, C., {Romano}, D., \& {Matteucci}, F. 2003, MNRAS, 339, 63

\bibitem[{{Daflon} {et~al.}(2009){Daflon}, {Cunha}, {de la Reza}, {Holtzman},
  \& {Chiappini}}]{daflonetal09}
{Daflon}, S., {Cunha}, K., {de la Reza}, R., {Holtzman}, J., \& {Chiappini}, C.
  2009, AJ, 138, 1577

\bibitem[{{Delgado-Inglada} {et~al.}(2014){Delgado-Inglada}, {Morisset}, \&
  {Stasi{\'n}ska}}]{delgadoingladaetal14}
{Delgado-Inglada}, G., {Morisset}, C., \& {Stasi{\'n}ska}, G. 2014, MNRAS, in
  press, arXiv:1402.4852

\bibitem[{{Esteban} {et~al.}(2009){Esteban}, {Bresolin}, {Peimbert},
  {Garc{\'{\i}}a-Rojas}, {Peimbert}, \& {Mesa-Delgado}}]{estebanetal09}
{Esteban}, C., {Bresolin}, F., {Peimbert}, M., {et~al.} 2009, ApJ, 700, 654

\bibitem[{{Esteban} {et~al.}(2013){Esteban}, {Carigi}, {Copetti},
  {Garc{\'{\i}}a-Rojas}, {Mesa-Delgado}, {Casta{\~n}eda}, \&
  {P{\'e}quignot}}]{estebanetal13}
{Esteban}, C., {Carigi}, L., {Copetti}, M.~V.~F., {et~al.} 2013, MNRAS, 433,
  382

\bibitem[{{Esteban} {et~al.}(2014){Esteban}, {Garc{\'{\i}}a-Rojas}, {Carigi}, {Peimbert},
  {Bresolin}, {L\'opez-S\'anchez}, \& {Mesa-Delgado}}]{estebanetal14}
{Esteban}, C., {Garc\'{\i}a-Rojas}, J. {Carigi}, L., {et~al.} 2014, MNRAS, 443,
  624

\bibitem[{{Esteban} {et~al.}(2005){Esteban}, {Garc{\'{\i}}a-Rojas}, {Peimbert},
  {Peimbert}, {Ruiz}, {Rodr{\'{\i}}guez}, \& {Carigi}}]{estebanetal05}
{Esteban}, C., {Garc{\'{\i}}a-Rojas}, J., {Peimbert}, M., {et~al.} 2005, ApJ,
  618, L95

\bibitem[{{Esteban} {et~al.}(2004){Esteban}, {Peimbert}, {Garc{\'{\i}}a-Rojas},
  {Ruiz}, {Peimbert}, \& {Rodr{\'{\i}}guez}}]{estebanetal04}
{Esteban}, C., {Peimbert}, M., {Garc{\'{\i}}a-Rojas}, J., {et~al.} 2004, MNRAS,
  355, 229

\bibitem[{{Esteban} {et~al.}(1998){Esteban}, {Peimbert}, {Torres-Peimbert}, \&
  {Escalante}}]{estebanetal98}
{Esteban}, C., {Peimbert}, M., {Torres-Peimbert}, S., \& {Escalante}, V. 1998,
  MNRAS, 295, 401

\bibitem[{{Esteban} {et~al.}(2002){Esteban}, {Peimbert}, {Torres-Peimbert}, \&
  {Rodr{\'{\i}}guez}}]{estebanetal02}
{Esteban}, C., {Peimbert}, M., {Torres-Peimbert}, S., \& {Rodr{\'{\i}}guez}, M.
  2002, ApJ, 581, 241

\bibitem[{{Ferland}(1999)}]{ferland99}
{Ferland}, G.~J. 1999, PASP, 111, 1524

\bibitem[{{Gail} \& {Sedlmayr}(1986)}]{gailsedlmayr86}
{Gail}, H.-P. \& {Sedlmayr}, E. 1986, A\&A, 166, 225

\bibitem[{{Galav\'{\i}s} {et~al.}(1995){Galav\'{\i}s}, {Mendoza}, \&
  {Zeippen}}]{galavisetal95}
{Galav\'{\i}s}, M.~E., {Mendoza}, C., \& {Zeippen}, C.~J. 1995, A\&AS, 111, 347

\bibitem[{{Galavis} {et~al.}(1997){Galavis}, {Mendoza}, \&
  {Zeippen}}]{galavisetal97}
{Galavis}, M.~E., {Mendoza}, C., \& {Zeippen}, C.~J. 1997, A\&AS, 123, 159

\bibitem[{{Garc{\'{\i}}a-Rojas} \& {Esteban}(2007)}]{garciarojasesteban07}
{Garc{\'{\i}}a-Rojas}, J. \& {Esteban}, C. 2007, ApJ, 670, 457

\bibitem[{{Harvey} {et~al.}(2008){Harvey}, {Huard}, {J{\o}rgensen},
  {Gutermuth}, {Mamajek}, {Bourke}, {Mer{\'{\i}}n}, {Cieza}, {Brooke},
  {Chapman}, {Alcal{\'a}}, {Allen}, {Evans}, {Di Francesco}, \&
  {Kirk}}]{harveyetal08}
{Harvey}, P.~M., {Huard}, T.~L., {J{\o}rgensen}, J.~K., {et~al.} 2008, ApJ,
  680, 495

\bibitem[{{Henry} {et~al.}(2000){Henry}, {Edmunds}, \&
  {K{\"o}ppen}}]{henryetal00}
{Henry}, R.~B.~C., {Edmunds}, M.~G., \& {K{\"o}ppen}, J. 2000, ApJ, 541, 660

\bibitem[{{Herbig} \& {Dahm}(2002)}]{herbigdahm02}
{Herbig}, G.~H. \& {Dahm}, S.~E. 2002, AJ, 123, 304

\bibitem[Herrero et al.(1992)]{herrero92} 
Herrero, A., Kudritzki, R.~P., Vilchez, J.~M., et al.\ 1992, \aap, 261, 209 

\bibitem[Hunter et al.(2008)]{hunter08} 
Hunter, I., Brott, I., Lennon, D.~J., et al.\ 2008, \apjl, 676, L29 


\bibitem[{{Irrgang} {et~al.}(2014){Irrgang}, {Przybilla}, {Heber}, {B{\"o}ck},
  {Hanke}, {Nieva}, \& {Butler}}]{irrgangetal14}
{Irrgang}, A., {Przybilla}, N., {Heber}, U., {et~al.} 2014, A\&A, in press,
  arXiv:1403.1122

\bibitem[Jaschek \& Gomez(1998)]{jaschek98} 
Jaschek, C., \& Gomez, A.~E.\ 1998, \aap, 330, 619 

\bibitem[{{Jenkins}(2009)}]{jenkins09}
{Jenkins}, E.~B. 2009, ApJ, 700, 1299

\bibitem[{{Jenkins}(2014)}]{jenkins14}
{Jenkins}, E.~B. 2014, in Life Cycle of Dust in the Universe, Observations,
  Theory and Laboratory Experiments

\bibitem[{{Kaufman} \& {Sugar}(1986)}]{kaufmansugar86}
{Kaufman}, V. \& {Sugar}, J. 1986, Journal of Physical and Chemical Reference
  Data, 15, 321

\bibitem[{{Keenan} {et~al.}(1993){Keenan}, {Hibbert}, {Ojha},
  \& {Caylon}}]{keenanetal93}
{Keenan}, F.~P., {Hibbert}, A., {Ojha}, P.~C., \& {Conlonl}, E.~. 1993, Phys. Scr., 48, 129

\bibitem[{{L\'opez-S\'anchez} {et~al.}(2007){L\'opez-S\'anchez}, {Esteban},
  {Garc{\'{\i}}a-Rojas}, {Peimbert}, \&
  {Rodr{\'{\i}}guez}}]{lopezsanchezetal07}
{L\'opez-S\'anchez}, A.~R., {Esteban}, C., {Garc{\'{\i}}a-Rojas}, J.,
  {Peimbert}, M., \& {Rodr{\'{\i}}guez}, M. 2007, ApJ, 656, 168

\bibitem[{{Luridiana} {et~al.}(2011){Luridiana}, {Garc\'{\i}a-Rojas}, {Aggarwal},  
{Bautista}, {Bergemann}, {Delahaye}, {del Zanna}, 
{Ferland}, {Lind}, {Manchado}, {Mendoza},  
{Mesa Delgado}, {N{\'u}{\~n}ez D{\'{\i}}az}, {Shaw}, \& 
{Wesson}}]{luridianaetal11}
{Luridiana}, V., {Garc\'{\i}a-Rojas}, J., {Aggarwal}, K., {et~al.} 2011, Summary of the Workshop: Uncertainties in Atomic Data and How They Propagate in Chemical Abundances, 
arXiv1110.1873

\bibitem[{{Luridiana} \& {Garc\'{\i}a-Rojas}(2012){Luridiana},\& {Garc\'{\i}a-Rojas}}]{luridianagarciarojas12}
{Luridiana}, V., \& {Garc\'{\i}a-Rojas}, J. 2012, in IAU Symposium, Vol.
  283, eds. A.~{Manchado}, L.~{Stanghellini}, 
  and D. {Sch{\"o}nberner}, 139--143

\bibitem[{{Luridiana} {et~al.}(2012){Luridiana}, {Morisset}, \&
  {Shaw}}]{luridianaetal12}
{Luridiana}, V., {Morisset}, C., \& {Shaw}, R.~A. 2012, in IAU Symposium, Vol.
  283, eds. A.~{Manchado}, L.~{Stanghellini}, 
  and D. {Sch{\"o}nberner}, 422--423

\bibitem[{{Luridiana} {et~al.}(2009){Luridiana}, {Sim{\'o}n-D{\'{\i}}az},
  {Cervi{\~n}o}, {Gonz{\'a}lez Delgado}, {Porter}, \&
  {Ferland}}]{luridianaetal09}
{Luridiana}, V., {Sim{\'o}n-D{\'{\i}}az}, S., {Cervi{\~n}o}, M., {et~al.} 2009,
  ApJ, 691, 1712

\bibitem[{{Mendoza} \& {Zeippen}(1982)}]{mendozazeippen82b}
{Mendoza}, C. \& {Zeippen}, C.~J. 1982, MNRAS, 199, 1025

\bibitem[{{Mesa-Delgado} {et~al.}(2009{\natexlab{a}}){Mesa-Delgado}, {Esteban},
  {Garc{\'{\i}}a-Rojas}, {Luridiana}, {Bautista}, {Rodr{\'{\i}}guez},
  {L{\'o}pez-Mart{\'{\i}}n}, \& {Peimbert}}]{mesadelgadoetal09b}
{Mesa-Delgado}, A., {Esteban}, C., {Garc{\'{\i}}a-Rojas}, J., {et~al.}
  2009{\natexlab{a}}, MNRAS, 395, 855

\bibitem[{{Mesa-Delgado} {et~al.}(2009{\natexlab{b}}){Mesa-Delgado},
  {L{\'o}pez-Mart{\'{\i}}n}, {Esteban}, {Garc{\'{\i}}a-Rojas}, \&
  {Luridiana}}]{mesadelgadoetal09}
{Mesa-Delgado}, A., {L{\'o}pez-Mart{\'{\i}}n}, L., {Esteban}, C.,
  {Garc{\'{\i}}a-Rojas}, J., \& {Luridiana}, V. 2009{\natexlab{b}}, MNRAS, 394,
  693

\bibitem[{{Mesa-Delgado} {et~al.}(2012){Mesa-Delgado},
  {N{\'u}{\~n}ez-D{\'{\i}}az}, {Esteban}, {Garc{\'{\i}}a-Rojas},
  {Flores-Fajardo}, {L{\'o}pez-Mart{\'{\i}}n}, {Tsamis}, \&
  {Henney}}]{mesadelgadoetal12}
{Mesa-Delgado}, A., {N{\'u}{\~n}ez-D{\'{\i}}az}, M., {Esteban}, C., {et~al.}
  2012, MNRAS, 426, 614

 \bibitem[Morel et al.(2006)]{morel06} 
Morel, T., Butler, K., Aerts, C., Neiner, C., \& Briquet, M.\ 2006, \aap, 457, 651 

\bibitem[{{Nicholls} {et~al.}(2012){Nicholls}, {Dopita}, \&
  {Sutherland}}]{nichollsetal12}
{Nicholls}, D.~C., {Dopita}, M.~A., \& {Sutherland}, R.~S. 2012, ApJ, 752, 148

\bibitem[{{Nicholls} {et~al.}(2013){Nicholls}, {Dopita}, {Sutherland},
  {Kewley}, \& {Palay}}]{nichollsetal13}
{Nicholls}, D.~C., {Dopita}, M.~A., {Sutherland}, R.~S., {Kewley}, L.~J., \&
  {Palay}, E. 2013, ApJS, 207, 21

\bibitem[{{Nieva} \& {Przybilla}(2012)}]{nievaprzybilla12}
{Nieva}, M.-F. \& {Przybilla}, N. 2012, A\&A, 539, A143

\bibitem[{{Nieva} \& {Sim{\'o}n-D{\'{\i}}az}(2011)}]{nievasimondiaz11}
{Nieva}, M.-F. \& {Sim{\'o}n-D{\'{\i}}az}, S. 2011, A\&A, 532, A2

\bibitem[{{Pe{\~n}a-Guerrero} {et~al.}(2012){Pe{\~n}a-Guerrero}, {Peimbert}, \&
  {Peimbert}}]{penaguerreroetal12b}
{Pe{\~n}a-Guerrero}, M.~A., {Peimbert}, A., \& {Peimbert}, M. 2012, ApJ, 756,
  L14

\bibitem[{{Peimbert}(2003)}]{apeimbert03}
{Peimbert}, A. 2003, ApJ, 584, 735

\bibitem[{{Peimbert}(1967)}]{peimbert67}
{Peimbert}, M. 1967, ApJ, 150, 825

\bibitem[{{Peimbert}(2008)}]{peimbert08}
{Peimbert}, M. 2008, Current Science, 95, 1165

\bibitem[{{Peimbert} \& {Costero}(1969)}]{peimbertcostero69}
{Peimbert}, M. \& {Costero}, R. 1969, Boletin de los Observatorios de Tonantzintla y Tacubaya, 5, 3

\bibitem[{{Peimbert} \& {Peimbert}(2013)}]{peimbertpeimbert13} 
Peimbert, A., \& Peimbert, M.\ 2013, ApJ, 778, 89 

\bibitem[{{Peimbert} {et~al.}(2007){Peimbert}, {Peimbert}, {Esteban},
  {Garc{\'{\i}}a-Rojas}, {Bresolin}, {Carigi}, {Ruiz}, \&
  {L{\'o}pez-S{\'a}nchez}}]{peimbertetal07}
{Peimbert}, M., {Peimbert}, A., {Esteban}, C., {et~al.} 2007, in Revista
  Mexicana de Astronomia y Astrofisica, vol. 27, Vol.~29, Revista Mexicana de
  Astronomia y Astrofisica Conference Series, ed. R.~{Guzm{\'a}n}, 72--79

\bibitem[{{Peimbert} {et~al.}(1993){Peimbert}, {Storey}, \&
  {Torres-Peimbert}}]{peimbertetal93}
{Peimbert}, M., {Storey}, P.~J., \& {Torres-Peimbert}, S. 1993, ApJ, 414, 626

\bibitem[{{Podobedova} {et~al.}(2009){Podobedova}, {Kelleher}, \&
  {Wiese}}]{podobedovaetal09}
{Podobedova}, L.~I., {Kelleher}, D.~E., \& {Wiese}, W.~L. 2009, Journal of
  Physical and Chemical Reference Data, 38, 171

\bibitem[{{Pradhan} {et~al.}(2006){Pradhan}, {Montenegro}, {Nahar}, \&
  {Eissner}}]{pradhanetal06}
{Pradhan}, A.~K., {Montenegro}, M., {Nahar}, S.~N., \& {Eissner}, W. 2006,
  MNRAS, 366, L6

\bibitem[{{Puls} {et~al.}(1996){Puls}, {Kudritzki}, {Herrero}, {Pauldrach},
  {Haser}, {Lennon}, {Gabler}, {Voels}, {V\'{\i}lchez}, {Wachter}, \& {Feldmeier}}]{pulsetal96}
{Puls}, J., {Kudritzki}, R.-P., {Herrero}, R., {et~al.} 1996, A\&A, 305, 171

\bibitem[{{Puls} {et~al.}(2005){Puls}, {Urbaneja}, {Venero}, {Repolust},
  {Springmann}, {Jokuthy}, \& {Mokiem}}]{pulsetal05}
{Puls}, J., {Urbaneja}, M.~A., {Venero}, R., {et~al.} 2005, A\&A, 435, 669

\bibitem[{{Ramsbottom} {et~al.}(1996){Ramsbottom}, {Bell}, \&
  {Stafford}}]{ramsbottometal96}
{Ramsbottom}, C.~A., {Bell}, K.~L., \& {Stafford}, R.~P. 1996, Atomic Data and
  Nuclear Data Tables, 63, 57

\bibitem[{{Rivero Gonz{\'a}lez} {et~al.}(2011){Rivero Gonz{\'a}lez}, {Puls}, \&
  {Najarro}}]{riverogonzalezetal11}
{Rivero Gonz{\'a}lez}, J.~G., {Puls}, J., \& {Najarro}, F. 2011, A\&A, 536, A58

\bibitem[{{Rivero Gonz{\'a}lez} {et~al.}(2012){Rivero Gonz{\'a}lez}, {Puls},
  {Najarro}, \& {Brott}}]{riverogonzalezetal12}
{Rivero Gonz{\'a}lez}, J.~G., {Puls}, J., {Najarro}, F., \& {Brott}, I. 2012,
  A\&A, 537, A79

\bibitem[{{Santolaya-Rey} {et~al.}(1997){Santolaya-Rey}, {Puls}, \&
  {Herrero}}]{santolayareyetal97}
{Santolaya-Rey}, A.~E., {Puls}, J., \& {Herrero}, A. 1997, A\&A, 323, 488

\bibitem[{{Sim{\'o}n-D{\'{\i}}az}(2010)}]{simondiaz10}
{Sim{\'o}n-D{\'{\i}}az}, S. 2010, A\&A, 510, A22

\bibitem[{{Sim{\'o}n-D{\'{\i}}az}
  {et~al.}(2011{\natexlab{a}}){Sim{\'o}n-D{\'{\i}}az}, {Castro}, {Herrero},
  {Puls}, {Garcia}, \& {Sab{\'{\i}}n-Sanjuli{\'a}n}}]{simondiazetal11b}
{Sim{\'o}n-D{\'{\i}}az}, S., {Castro}, N., {Herrero}, A., {et~al.}
  2011{\natexlab{a}}, Journal of Physics Conference Series, 328, 012021

\bibitem[{{Sim{\'o}n-D{\'{\i}}az}
  {et~al.}(2011{\natexlab{b}}){Sim{\'o}n-D{\'{\i}}az}, {Garc{\'{\i}}a-Rojas},
  {Esteban}, {Stasi{\'n}ska}, {L{\'o}pez-S{\'a}nchez}, \&
  {Morisset}}]{simondiazetal11}
{Sim{\'o}n-D{\'{\i}}az}, S., {Garc{\'{\i}}a-Rojas}, J., {Esteban}, C., {et~al.}
  2011{\natexlab{b}}, A\&A, 530, A57

\bibitem[{{Sim{\'o}n-D{\'{\i}}az} \& {Herrero}(2014)}]{simondiazherrero14}
{Sim{\'o}n-D{\'{\i}}az}, S. \& {Herrero}, A. 2014, A\&A, 562, A135

\bibitem[{{Sim{\'o}n-D{\'{\i}}az} \&
  {Stasi{\'n}ska}(2011)}]{simondiazstasinska11}
{Sim{\'o}n-D{\'{\i}}az}, S. \& {Stasi{\'n}ska}, G. 2011, A\&A, 526, A48

\bibitem[{{Sofia} {et~al.}(1994){Sofia}, {Cardelli}, \& {Savage}}]{sofiaetal94}
{Sofia}, U.~J., {Cardelli}, J.~A., \& {Savage}, B.~D. 1994, ApJ, 430, 650

\bibitem[{{Sofia} \& {Meyer}(2001)}]{sofiameyer01}
{Sofia}, U.~J., \& {Meyer}, D.~M. 2001, ApJ, 554, L221

\bibitem[{{Stasi{\'n}ska} {et~al.}(2007){Stasi{\'n}ska}, {Tenorio-Tagle},
  {Rodr\'{\i}guez}, \& {Henney}}]{stasinskaetal07}
{Stasi{\'n}ska}, G., {Tenorio-Tagle}, G., {Rodr\'{\i}guez}, M., \& {Henney},
  W.~J. 2007, A\&A, 471, 193

\bibitem[{Storey} \& {Hummer}(1995)]{storeyhummer95}
{Storey}, P.~J., \& {Hummer}, D.~G.\ 1995, MNRAS, 272, 41

\bibitem[{{Tayal}(2011)}]{tayal11}
{Tayal}, S.~S. 2011, ApJS, 195, 12

\bibitem[{{Tayal} \& {Gupta}(1999)}]{tayalgupta99}
{Tayal}, S.~S. \& {Gupta}, G.~P. 1999, ApJ, 526, 544

\bibitem[{{Telting} {et~al.}(2014)}]{telting14}
{Telting}, J.~H., {Avila}, G., {Buchhave}, L., {et~al.} 2014, AN, 335, 41

\bibitem[{{Torres-Peimbert} \& {Peimbert}(1977)}]{torrespeimbertpeimbert77}
{Torres-Peimbert}, S. \& {Peimbert}, M. 1977, Rev. Mexicana Astron. Astrofis.,
  2, 181

\bibitem[{{Tremonti} {et~al.}(2004){Tremonti}, {Heckman}, {Kauffmann},
  {Brinchmann}, {Charlot}, {White}, {Seibert}, {Peng}, {Schlegel}, {Uomoto},
  {Fukugita}, \& {Brinkmann}}]{tremontietal04}
{Tremonti}, C.~A., {Heckman}, T.~M., {Kauffmann}, G., {et~al.} 2004, ApJ, 613,
  898

\bibitem[{{Trundle} {et~al.}(2002){Trundle}, {Dufton}, {Lennon}, {Smartt}, \&
  {Urbaneja}}]{trundleetal02}
{Trundle}, C., {Dufton}, P.~L., {Lennon}, D.~J., {Smartt}, S.~J., \&
  {Urbaneja}, M.~A. 2002, A\&A, 395, 519

\bibitem[{{Tsamis} \& {P{\'e}quignot}(2005)}]{tsamispequignot05}
{Tsamis}, Y.~G. \& {P{\'e}quignot}, D. 2005, MNRAS, 364, 687

\bibitem[{{Tsamis} {et~al.}(2011){Tsamis}, {Walsh}, {V{\'{\i}}lchez}, \&
  {P{\'e}quignot}}]{tsamisetal11}
{Tsamis}, Y.~G., {Walsh}, J.~R., {V{\'{\i}}lchez}, J.~M., \& {P{\'e}quignot},
  D. 2011, MNRAS, 412, 1367

\bibitem[{{U} {et~al.}(2009){U}, {Urbaneja}, {Kudritzki}, {Jacobs}, {Bresolin},
  \& {Przybilla}}]{uetal09}
{U}, V., {Urbaneja}, M.~A., {Kudritzki}, R.-P., {et~al.} 2009, ApJ, 704, 1120

\bibitem[{{Wyse}(1942)}]{wyse42}
{Wyse}, A.~B. 1942, ApJ, 95, 356

\bibitem[{{White} \& {Sofia}(2011)}]{whitesofia11}
{White}, B. \& {Sofia}, U.~J. 2011, in American Astronomical Society Meeting
  Abstracts \#218, \#129.23

\bibitem[{{Zeippen}(1982)}]{zeippen82}
{Zeippen}, C.~J. 1982, MNRAS, 198, 111





\end{thebibliography}
%
%

\appendix

\section{On the distance to the Cocoon nebula as determined from BD~+46\,3474}\label{ap1}

There have been several independent determinations of the distance to the IC~5146
star-forming region, where the Cocoon nebula is located. The proposed values ranges
from 460 to 1400~pc. We refer the reader to \cite{harveyetal08} for a complete 
compilation of published distance estimates previous to 2008, and a detailed 
discussion of the various considered methodologies and their reliability. 

In this paper, we were mainly interested in the quantitative spectroscopic analysis
of BD+46~3474 to determine its photospheric chemical composition and compare the
derived abundances with those resulting from the study of the Cocoon nebula spectrum.
However, as a plus, we can also reevaluate the issue of the distance to this
star and its associated \hii\ region using state-of-the-art information. Below,
we describe the methodology we have followed and our proposed value.

Table~\ref{t3} summarized the spectroscopic parameters (\Teff  and \grav, among others)
resulting from the \fastwind\ analysis, as well as some photometric information that
we use for the evaluation of the distance (namely, the V magnitude and the B-V color).
From the comparsion of intrinsic (B-V)$_0$ color predicted by a \fastwind\ model 
with the indicated \Teff\ and \grav\ and the observed value, we obtained the
value of the extinction parameter in the V band. We hence determined $M_v$, $R$, log$L$,
and $M_{\rm sp}$ (spectroscopic mass) assuming several values of the distance. The absolute 
visual magnitude was computed by means of
\begin{equation}
 M_v=V-5logd+5-A_v
\end{equation}
and the stellar radius, luminosity, and spectroscopic mass was derived by means of the
strategy indicated in \cite{herrero92}. Last, we located the star in the HR diagram
and computed the evolutionary mass by comparing with the evolutionary tracks by 
\cite{brott11}.

Table~\ref{tA} and Figure~\ref{fA} summarizes the results from this exercise. From
inspection of Figure~\ref{fA} it becomes clear that any distance below 720~pc is
not possible since the star would be located below the zero-age main sequence (ZAMS)
line. We have selected four distances above that value. The largest one (1.2 Kpc)
is the value proposed by \cite{herbigdahm02} based on the spectroscopic distances to
the late-B stars and two different main-sequence calibrations: \cite{jaschek98} 
absolute magnitudes for B dwarf standards, and the Schmidt-Kaler ZAMS 
\cite{aller82}. The other values are our suggested distance (800 pc), the value
proposed by \cite{harveyetal08}, and an intermediate value.

As illustrated in Table{\ref{tA} and Figure~\ref{fA} the determined spectroscopic
mass, evolutionary mass and age are a function of the assumed distance. We can clearly
discard the 1200~pc (and even the 950~pc) solutions since these distances result in
a very bad agreement between the spectroscopic and evolutionary masses and a too
evolved star ($\ge$~4 Myr). Note that given the high number of accreting pre-MS stars,
we expect the age of the IC~5146 cluster to be less than a few Myr 
\citep{herbigdahm02,harveyetal08}. We hence use the $M_{\rm sp}$=$M_{\rm ev}$ criterium
to propose 800$\pm$80~pc as the distance to BD+46~3474.

%
\begin{table*}[t!]
\begin{center}
\caption{\footnotesize Computed values of absolute visual magnitude, stellar
radius, luminosity, spectroscopic mass, evolutionary mass and age for different
assumed distances to BD+46~3474 (see Sect.~\ref{ap1} for explanations). We indicate in bold our proposed distance.}
\label{tA}
\begin{tabular}{c | c c c c c c}    
\noalign{\smallskip}
\hline
\hline
\noalign{\smallskip}
d (pc) & M$_{\rm v}$ & R (R$_{\odot}$) & log(L/L$_{\odot}$) & M$_{\rm sp}$ (M$_{\odot}$) & M$_{\rm ev}$ (M$_{\odot}$) & Age (Myr) \\ 
\noalign{\smallskip}
\hline
\noalign{\smallskip}
720 & -2.8 & 4.7 & 4.23 & 13 & 14--15 & ZAMS \\
{\bf 800} & {\bf -3.0} & {\bf 5.2} & {\bf 4.31 } & {\bf 15 } & {\bf $\sim$15} & {\bf 2} \\
850 & -3.2 & 5.5 & 4.37 & 18 & 15--16 & 3 \\
950 & -3.4 & 6.2 & 4.47 & 22 & $\sim$16 & 4 \\
1200 & -3.9 & 7.8 & 4.68 & 35 & $\sim$18 & 6\\
\noalign{\smallskip}
\hline
\hline
\end{tabular}
\end{center}
\end{table*}

%
\begin{figure}[!ht]
\centering
\includegraphics[width=8.6cm,angle=0]{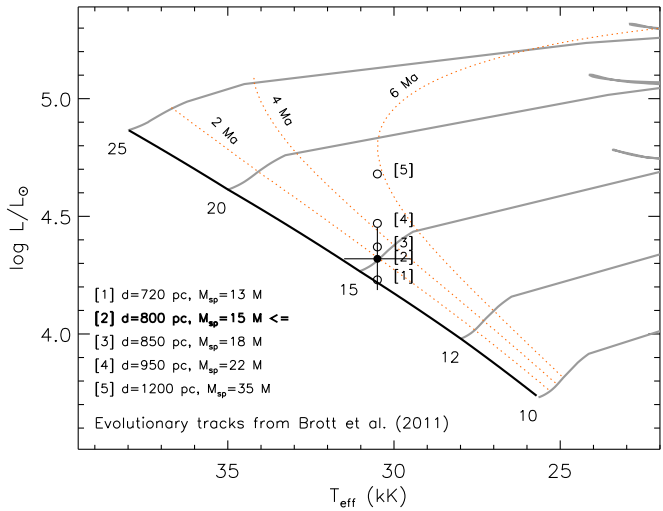}
\caption{Location of BD+46~3474 in the HR diagram for different values of assumed
distance (see also Table~\ref{tA}. Evolutionary tracks and isocrones from
\cite{brott11}. The corresponding spectroscopic masses are also indicated.}
\label{fA}
\end{figure}

%
\end{document}